\documentclass[12pt,draftclsnofoot,onecolumn,journal,letterpaper]{IEEEtran}
\usepackage{float}
\usepackage{amssymb}
\usepackage{polynomial}
\usepackage{color}
\usepackage[dvipsnames]{xcolor}
\usepackage{algpseudocode}
\usepackage{cite}
\usepackage{graphicx}
\usepackage{esdiff}
\usepackage{stfloats}
\usepackage{amsthm}
\theoremstyle{plain}

\newtheorem{lemma}{Lemma}

\usepackage{amsfonts}
\usepackage[cmex10]{amsmath}
\ifCLASSOPTIONcompsoc
\usepackage[tight,normalsize,sf,SF]{subfigure}
\else
\usepackage[tight,footnotesize]{subfigure}
\fi

\usepackage{amsmath}
\usepackage{epstopdf}
\usepackage{bbm}
\usepackage{array}
\usepackage{mdwmath}
\usepackage{mdwtab}
\usepackage{multirow}
\usepackage{hyperref}
\usepackage{algorithm}
\usepackage{tabularx, booktabs}
\newcolumntype{Y}{>{\centering\arraybackslash}X}

\begin{document}

\title{Modeling the Time-varying Subjective Quality of HTTP Video Streams with Rate Adaptations}

\author{Chao Chen,~\IEEEmembership{Student Member,~IEEE,}
        Lark Kwon Choi,~\IEEEmembership{Student Member,~IEEE,}
        Gustavo~de~Veciana,~\IEEEmembership{Fellow,~IEEE,}
        Constantine Caramanis,~\IEEEmembership{Member,~IEEE,}
        Robert W. Heath Jr.,~\IEEEmembership{Fellow,~IEEE,}
        and Alan C. Bovik,~\IEEEmembership{Fellow,~IEEE}

\thanks{The authors are with Department of Electrical and Computer Engineering,
The University of Texas at Austin, Austin, TX 78712-0240, USA. This research was supported in part by Intel Inc. and Cisco Corp. under the VAWN program.}
}


\maketitle

\begin{abstract}
Newly developed HTTP-based video streaming technologies enable flexible rate-adaptation under varying channel conditions. Accurately predicting the users' Quality of Experience (QoE) for rate-adaptive HTTP video streams is thus critical to achieve efficiency. An important aspect of understanding and modeling QoE is predicting the up-to-the-moment subjective quality of a video as it is played, which is difficult due to hysteresis effects and nonlinearities in human behavioral responses. This paper presents a Hammerstein-Wiener model for predicting the time-varying subjective quality (TVSQ) of rate-adaptive videos. To collect data for model parameterization and validation, a database of longer-duration videos with time-varying distortions was built and the TVSQs of the videos were measured in a large-scale subjective study. The proposed method is able to reliably predict the TVSQ of rate adaptive videos. Since the Hammerstein-Wiener model has a very simple structure, the proposed method is suitable for on-line TVSQ prediction in HTTP based streaming.
\end{abstract}


%
\IEEEpeerreviewmaketitle
\section{Introduction}
\label{sec:intro}
\IEEEPARstart{B}{ecause} the Hypertext Transfer Protocol (HTTP) is firewall-friendly, HTTP-based adaptive bitrate video streaming has become a popular alternative to its Real-Time Transport Protocol (RTP)-based counterparts. Indeed, companies such as Apple, Microsoft and Adobe have developed HTTP-based video streaming protocols \cite{smoothstream,livestream,dynamicsstream}, and the Moving Picture Experts Group (MPEG) has issued an international standard for HTTP based video streaming, called Dynamic Adaptive Streaming over HTTP (DASH) \cite{DASH}.

Another important motivation for HTTP-based adaptive bitrate video streaming is to reduce the risk of playback interruptions caused by channel throughput fluctuations. When a video is being transmitted, the received video data are first buffered at the receiver and then played out to the viewer. Since the channel throughput generally varies over time, the amount of buffered video decreases when the channel throughput falls below the video data rate. Once all the video data buffered at the receiver has been played out, the playback process stalls, significantly impacting the viewer's Quality of Experience (QoE) \cite{ZhangHui_QoE}\cite{zinner10}. In HTTP-based rate-adaptive streaming protocols, videos are encoded into multiple representations at different bitrates. Each representation is then partitioned into segments of lengths that are several seconds long. At any moment, the client can dynamically select a segment from an appropriate representation to download in order to adapt the downloading bitrate to its channel capacity. Although HTTP-based streaming protocols can effectively reduce the risk of playback interruptions, designing rate-adaptation methods that could optimize end-users' QoE is difficult since the relationship between the served bitrate and the users' viewing experience is not well understood. In particular, when the video bitrate is changed, the served video quality may also vary. If the impact of quality variations on QoE is not accurately predicted, the rate adaptation method will not provide the optimal QoE for the users.

One important indicator of QoE is the {\it time-varying subjective quality} (TVSQ) of the viewed videos. Assuming playback interruptions are avoided, the TVSQ is {\it a continuous-time record of viewers' judgments of the quality of the video as it is being played and viewed}. The TVSQ depends on many elements of the video including spatial distortions and temporal artifacts \cite{MOVIE,fluctuation}. What's more, human viewers exhibit a hysteresis \cite{SesBov11} or recency \cite{Pearson98} ``after effect", whereby the TVSQ of a video at a particular moment depends on the viewing experience before the moment. The quantitative nature of this dependency is critical for efficient rate adaptation. For example, as observed in our subjective study (see Section \ref{sec:pre_obs} for more detail), a viewer suffering a previous unpleasant viewing experience tends to penalize the perceived quality in the future. One approach to combat this is to force the rate controller to provide higher video quality in the future to counterbalance the negative impact of a prior poor viewing experience. But, without a predictive model for TVSQ, it is impossible to qualitatively assess how much quality improvement is needed. Another important property of the TVSQ is its nonlinearity. In particular, the sensitivity of the TVSQ to quality variation is not constant. This property should also be utilized for resource allocation among users sharing a network resource (such as transmission time in TDMA systems). For example, when the TVSQ of a user is insensitive to quality variations, the rate-controller could reserve some transmission resources by reducing the bitrate without lowering the user's TVSQ. The reserved resources could then be used to increase the bitrate of other users and thus improve their TVSQs. A predictive model for TVSQ is an essential tool to assess the sensitivity of TVSQ and to achieve quality-efficient rate adaptation.

The goal of this paper is to develop a predictive model that captures the impact of quality variations on TVSQ. The model predicts the average TVSQ every second and can be used to improve rate-adaptation algorithms for HTTP-based video streaming.

We propose to predict TVSQ in two steps. The two steps capture the spatial-temporal characteristics of the video and the hysteresis effects in human behavioral responses, respectively. In the first step, quality-varying videos are partitioned into one second long video chunks and the short-time subjective quality (STSQ) of each chunk is predicted. Unlike TVSQ, which is a temporal record, the STSQ is {\it a scalar prediction of viewers' subjective judgment of a {\it short video}'s overall perceptual quality}. A STSQ prediction model such as those in \cite{MOVIE,MSSSIM,VQM,ST-MAD,RRED} operates by extracting perceptually relevant spatial and temporal features from short videos then uses these to form predictions of STSQ. Hence, STSQ contains useful, but incomplete evidence about TVSQ. Here, the Video-RRED algorithm \cite{RRED} is employed to predict STSQs because of its excellent quality prediction performance and fast computational speed. In the second step, the predicted STSQs are sent to a dynamic system model, which predicts the average TVSQ every second. The model mimics the hysteresis effects with a linear filter and captures the nonlinearity in human behavior with nonlinear functions at the input and the output of the linear filter.
In HTTP-based streaming protocols, the interval between consecutive video data rate adaptations is usually several seconds long\footnote{For example, in MPEG-DASH \cite{DASH}, the rate adaptation interval is at least two seconds.}. Since the proposed model predicts the average TVSQ per second, the prediction timescales are suitable for HTTP-based streaming.


\begin{figure}[h]
\centering
\includegraphics[width=0.9\columnwidth]{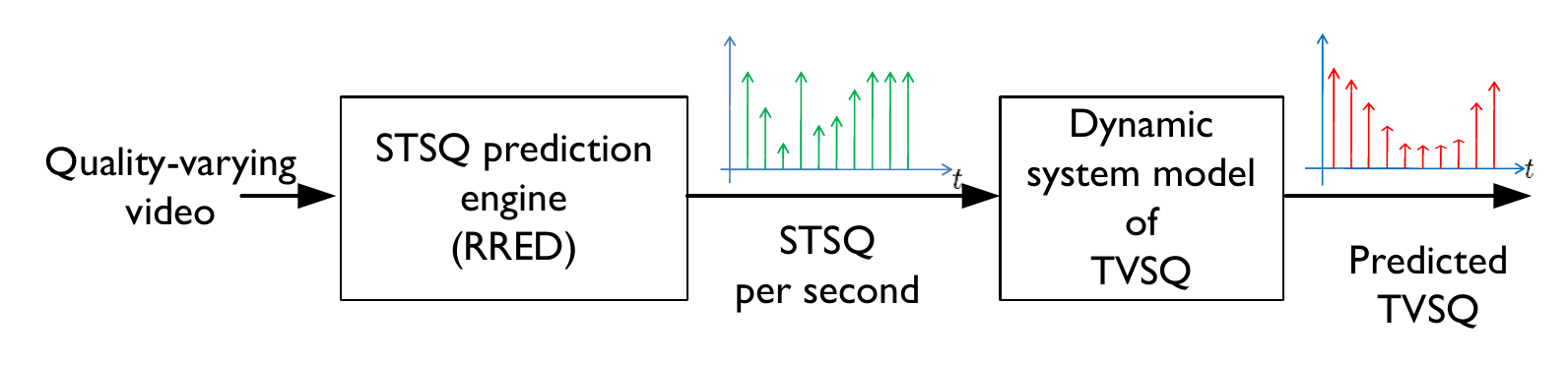}
\caption{\label{fig:pqmodel}Proposed paradigm for TVSQ prediction.}
\end{figure}

The contributions of this paper are summarized as follows:
\begin{enumerate}
\item{\it A new database for the TVSQ of HTTP-based video streams.} A database of rate-varying video sequences is built to simulate quality fluctuations commonly encountered in video streaming applications. \footnote{Since HTTP is based on TCP, which guarantees that the data is delivered without packet loss. Thus, only encoding distortions are considered.}. Then, a subjective study was conducted to measure the TVSQs of these video sequences. This database is useful for developing and validating TVSQ models and thus is important in its own right, as it may contribute to future research efforts.

\item{\it An effective TVSQ prediction method.} Using the new database, a dynamic system model is proposed to predict the average TVSQ per second of video. Experimental results show that the proposed model reliably tracks the TVSQ of video sequences with time-varying qualities. The dynamic system model has simple structure and is computationally efficient for TVSQ prediction. It is in fact suitable for online TVSQ-optimized rate adaptation. In HTTP-based video streaming protocols, the video is encoded into multiple representations at different video data rates. These representations are stored on the video server before transmission. Thus, the rate-STSQ function for each second of the video can be computed off-line before transmission. Since the proposed dynamic system model predicts the TVSQ from the STSQ, we may combine the rate-STSQ function with the dynamic system model to obtain a rate-TVSQ model. This rate-TVSQ model can then be used to determine the video data rate that optimizes the TVSQ.
\end{enumerate}

\noindent\textbf{Related Work}:
TVSQ is an important research subject in the realm of visual quality assessment \cite{Pearson98,Tan98,MasHem2004,Barkowsky07,SesBov11,NN}. In \cite{Pearson98}, the relationship between STSQ and TVSQ for packet videos transmitted over ATM networks was studied. A so-called ``recency effect" was observed in their subjective experiments. At any moment, the TVSQ is quite sensitive to the STSQs over the previous (at least) 20-30 seconds \cite{Pearson98}. Thus, the TVSQ at any moment depends not only on the current video quality, but also on the preceding viewing experience. In \cite{Tan98}, Tan {\it et al.}\ proposed an algorithm to estimate TVSQ. They first applied an image quality assessment algorithm to each video frame. Then they predicted the TVSQ with per-frame qualities using a ``cognitive emulator" designed to capture the hysteresis of the human behavioral responses to visual quality variations. The performance of this model was evaluated on a database of three videos, on which the encoding data rates were adapted over a slow time scale of 30-40 seconds \cite{Tan98}. In \cite{MasHem2004}, a first-order infinite impulse response (IIR) filter was used to predict the TVSQ based on per-frame distortions, which were predicted by spatial and temporal features extracted from the video. This method was shown to track the dynamics of the TVSQ on low bit-rate videos. In \cite{Barkowsky07}, an adaptive IIR filter was proposed to model the TVSQ. Since the main objective of \cite{Barkowsky07} was to predict the {\em overall} subjective quality of a long video sequence using the predicted TVSQ, the performance of this model was not validated against the measured TVSQ. In \cite{SesBov11}, a temporal pooling strategy was employed to map the STSQ to the overall video quality using a model of visual hysteresis. As an intermediate step, the STSQ was first mapped to the TVSQ, then the overall quality was estimated as a time-averaged TVSQ. Although this pooling strategy yields good predictions of the overall video quality, the model for the TVSQ is a non-causal system, which contradicts the fact that the TVSQ at a moment only depends on current and previous STSQs. In \cite{NN}, a convolutional neural network was employed to map features extracted from each video frame to the TVSQ. The estimated TVSQs were shown to achieve high correlations with the measured TVSQ values on constant bitrate videos.

In \cite{Barkowsky07} and \cite{SesBov11}, estimated TVSQ was used as an intermediate result in an overall video quality prediction process. However, the performances of these models were not validated against recorded subjective TVSQ. The TVSQ models proposed in \cite{MasHem2004}\cite{NN}\cite{Tan98} mainly targeted videos for which the encoding rate was fixed or changing slowly. Newly proposed HTTP-based video streaming protocols, e.g., DASH, provide the flexibility to adapt video bitrates over time-scales as short as 2 seconds. Thus the prior models cannot be directly applied to estimate the TVSQ for HTTP-based video streaming.


The advantages of our work are summarized as follows:
\begin{itemize}
  \item {\it The proposed TVSQ estimation method is designed for HTTP-based video streaming.} In this paper, a new video quality database is built and is specifically configured to enable the development of TVSQ prediction models of HTTP-based video streaming. The STSQs of the videos in the new database were designed to vary randomly over time scales of several seconds in order to simulate the quality variations encountered in HTTP-based video streaming. The database consists of 15 videos. Each video is 5 minutes long and is viewed by 25 subjects.
  \item {\it Temporal distortion is considered.} Previous TVSQ models measured the quality of individual video frames, then used a TVSQ emulator to estimate the TVSQ using the per-frame quality values \cite{Tan98}. However, neither per-frame quality nor TVSQ emulation captures temporal distortions such as mosquito effects, motion compensation mismatches or jerkiness. In our proposed method, temporal distortions such as these are captured by the Video-RRED STSQ predictor which is based on a natural scene statistical model of adjacent video frames.
\end{itemize}
\noindent\textbf{Organization and Notation:} The remainder of this paper is organized as follows: Section \ref{sec:study} introduces the new TVSQ database and describe its construction. In Section \ref{sec:model_id} explains the model for TVSQ prediction. In Section~\ref{sec:model_analysis}, the model is validated through extensive experimentation and by a detailed system theoretic analysis.

Some of the key notation are briefly introduced as follows. Let $\{x[t],~t=1,2\dots\}$ denote discrete time series. The notation $\big({x}\big)_{t_1:t_2}$ denotes the column vector $\left({x[t_1]}, x[t_1+1], \dots, {x[{t_2}]}\right)$. The zero-padded convolution of $\big({x}\big)_{t_1:t_2}$ and $\big({y}\big)_{t_1:t_2}$ is denoted by $\big({x}\big)_{t_1:t_2}*\big({y}\big)_{t_1:t_2}$. Lower-case symbols such as $a$ denote scalar variables. Random variables are denoted by uppercase letters such as $\mathsf{A}$. Boldface lower-case symbols such as $\mathbf a$ denote column vectors and $\mathbf{a}^\mathsf{T}$ is the transpose of $\mathbf a$. Calligraphic symbols such as $\mathcal{A}$ denote sets while $|\mathcal A|$ is the cardinality of $\mathcal{A}$. Finally, the function $\nabla_{\mathbf a}\mathrm{f}(\mathbf{a,b})$ denotes the gradient of the multivariate function $\mathrm f(\mathbf{a,b})$ with respect to variable $\mathbf{a}$.
\section{Subjective Study for Model Identification}
\label{sec:study}
In this section, the construction of the database and the design of the subjective experiments is described first. Then, based on the experimental results, the dynamic system model for TVSQ prediction is motivated.

\subsection{Quality-varying Video Construction}
\label{sec:database}

Using the following 5 steps, 15 quality-varying videos were constructed such that their STSQs vary randomly across time.

\newcounter{itemcounter}
\begin{list}
{\textbf{\arabic{itemcounter}.}}
{\usecounter{itemcounter}\leftmargin=1.4em}
\item Eight high quality, uncompressed video clips with different content were selected. These clips have a spatial resolution of 720p ($1280\times720$) and a frame rate of 30 fps. A short description of these clips is provided in Table~\ref{tab:video_descrpt}. The content was chosen to represent a broad spectrum of spatial and temporal complexity (see sample frames in Fig.~\ref{fig:sampleframe}).
\begin{table}
\centering
\caption{\label{tab:video_descrpt}A brief description of the video clips in our database.}
\begin{tabular}{|c|c|l|}
\hline
Name&Abbreviation&Description\\
\hline
Fountain&ft&Still camera, shows a fountain.\\
\hline
Turtles&tu& Still camera, a girl is feeding turtles.\\
\hline
Stick&st& Still camera, a man is waving a stick.\\
\hline
Bulldozer&bu&Camera span, a man is driving a bulldozer.\\
\hline
Singer\&girl&sg&Camera zoom, a man is singing to a girl.\\
\hline
Volleyball&vo&Still camera, shows a volleyball game.\\
\hline
Dogs&do&Camera span, two dogs play near a pool.\\
\hline
Singer&si&Camera zoom, a singer is singing a song.\\
\hline

\end{tabular}
\end{table}

\begin{figure}[!h]
\centering
\subfigure[ft]{
\includegraphics[scale=0.08]{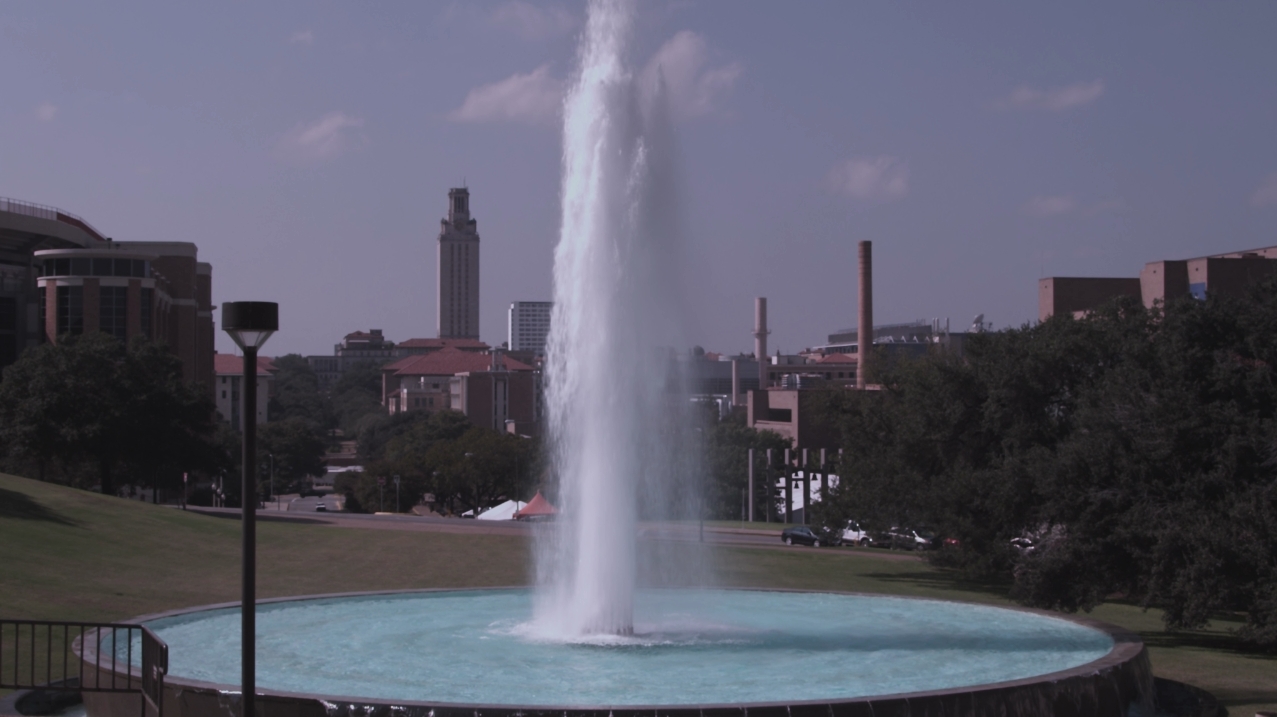}

}
\subfigure[tu]{
\includegraphics[scale=0.132]{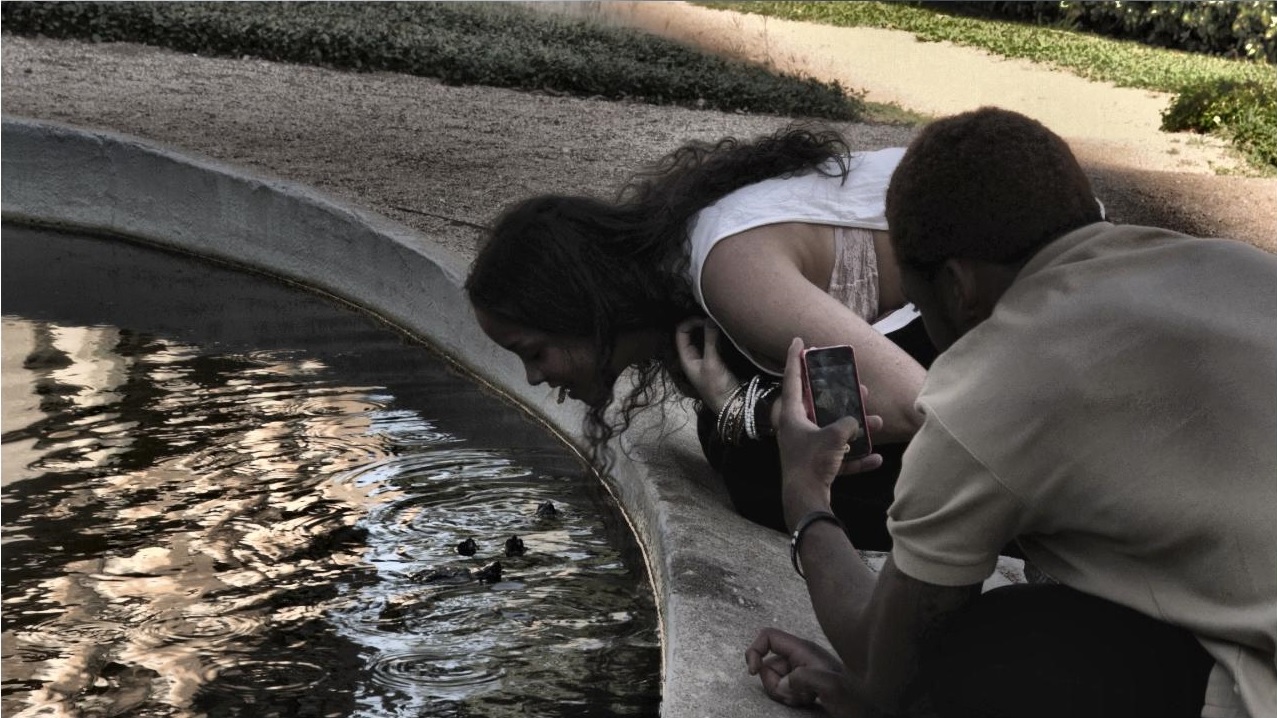}

}
\subfigure[st]{
\includegraphics[scale=0.132]{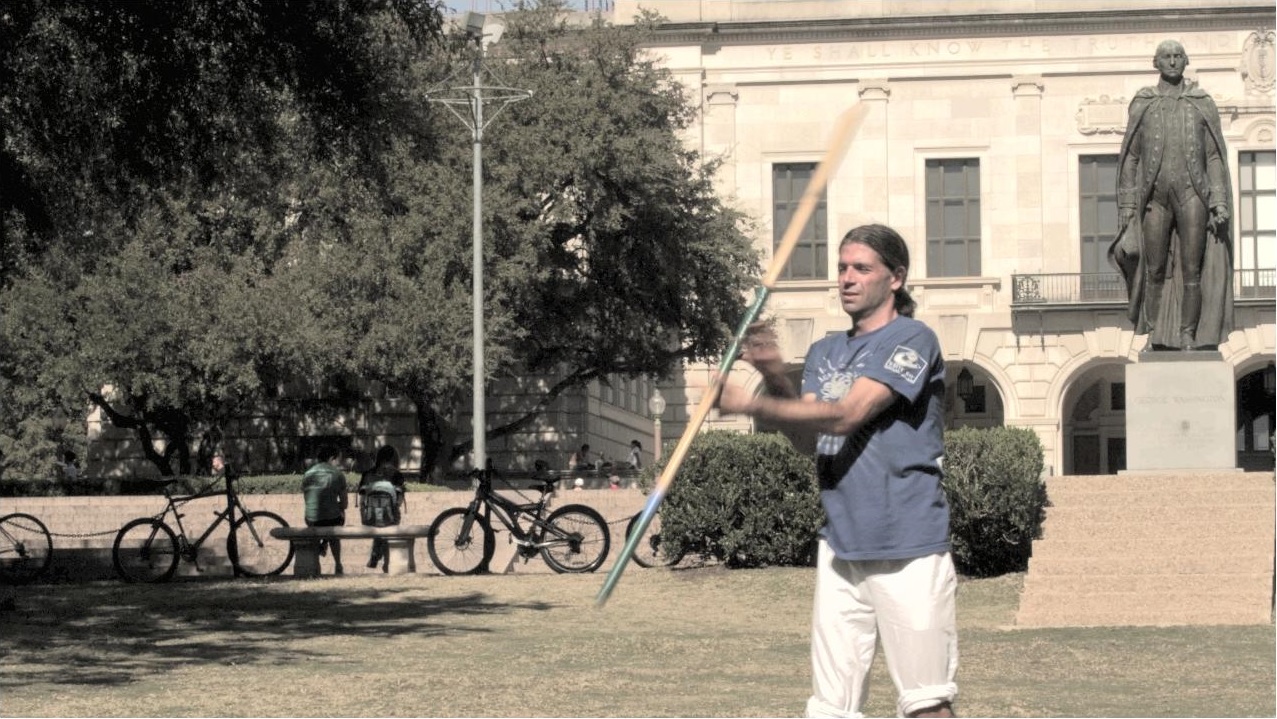}

}
\subfigure[bu]{
\includegraphics[scale=0.132]{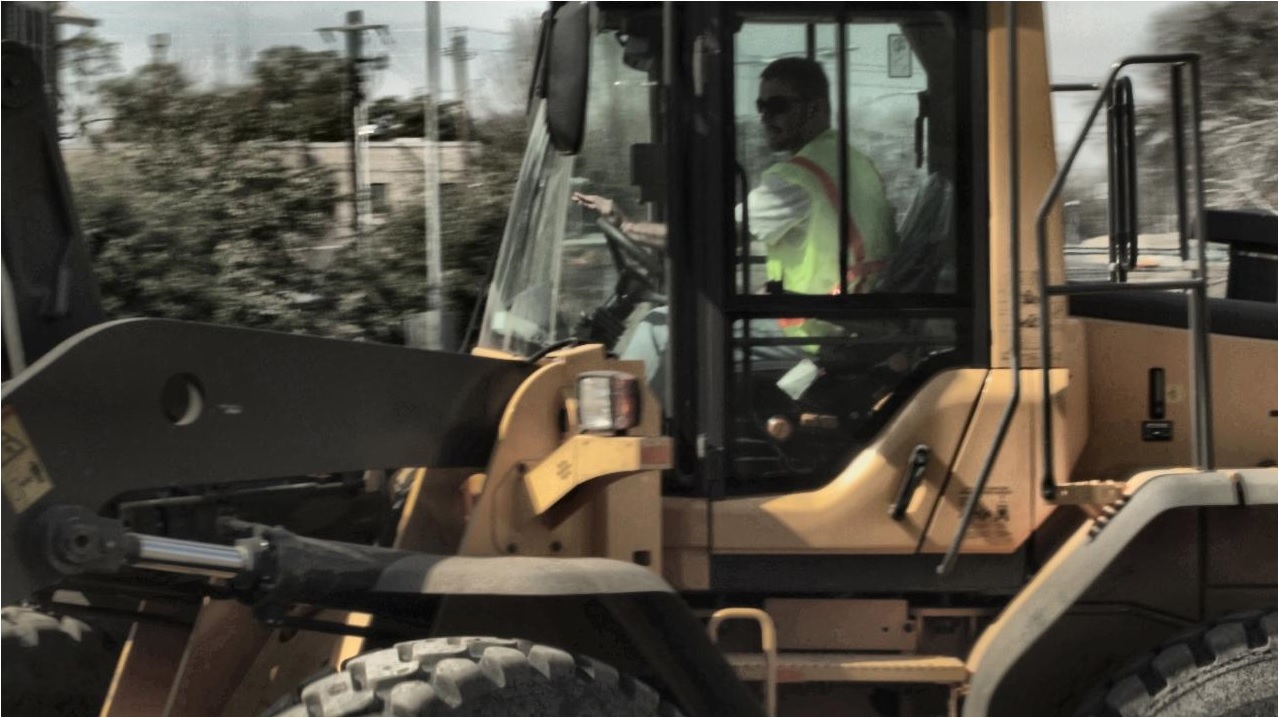}

}
\subfigure[sg]{
\includegraphics[scale=0.132]{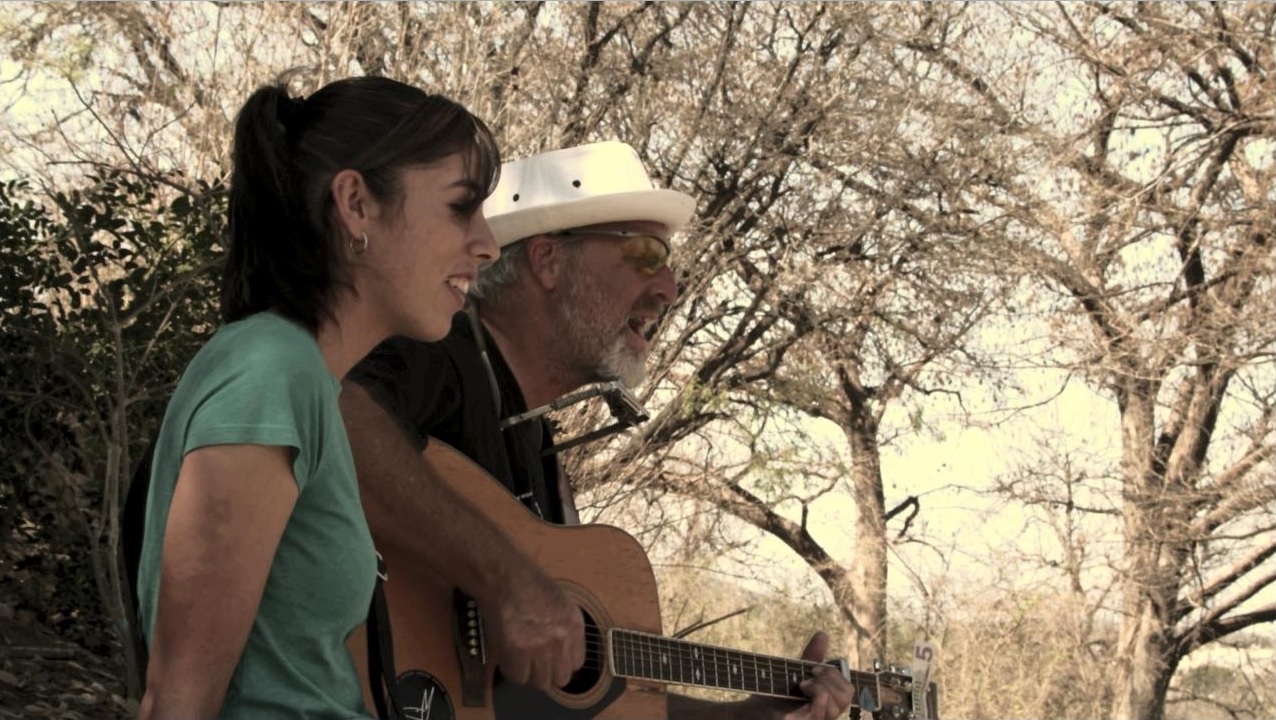}

}
\subfigure[vo]{
\includegraphics[scale=0.08]{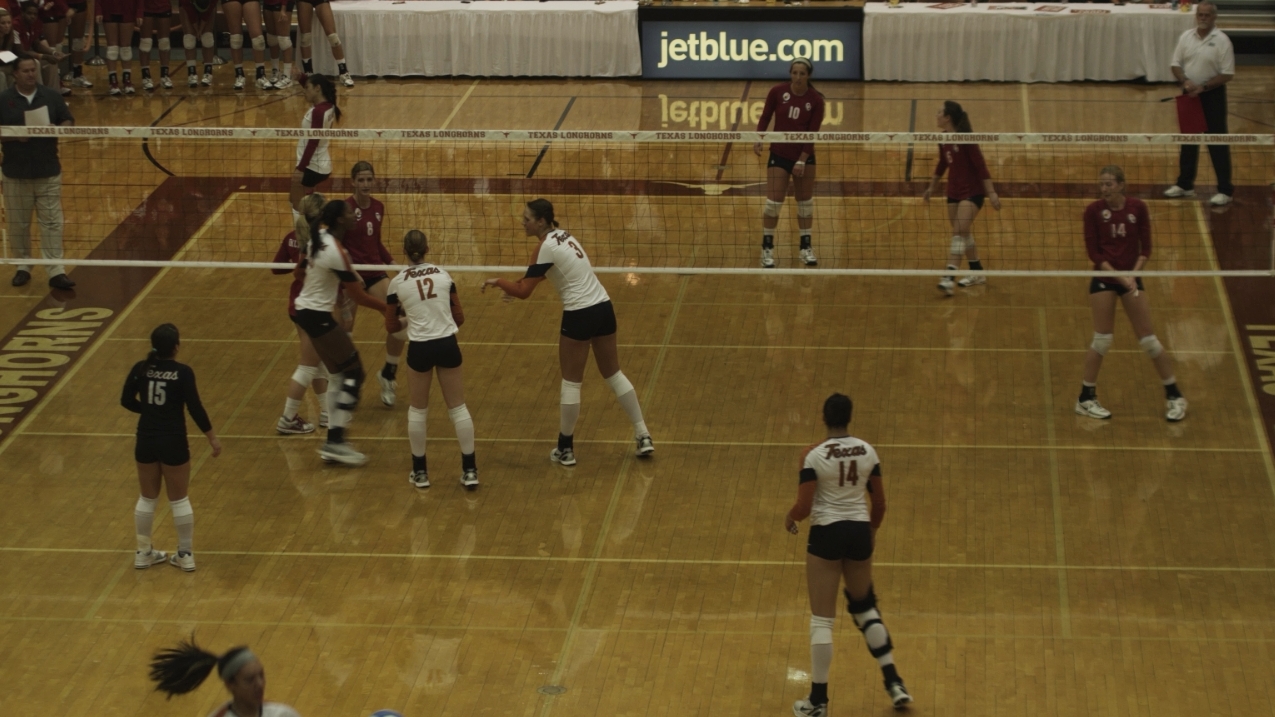}

}
\subfigure[do]{
\includegraphics[scale=0.08]{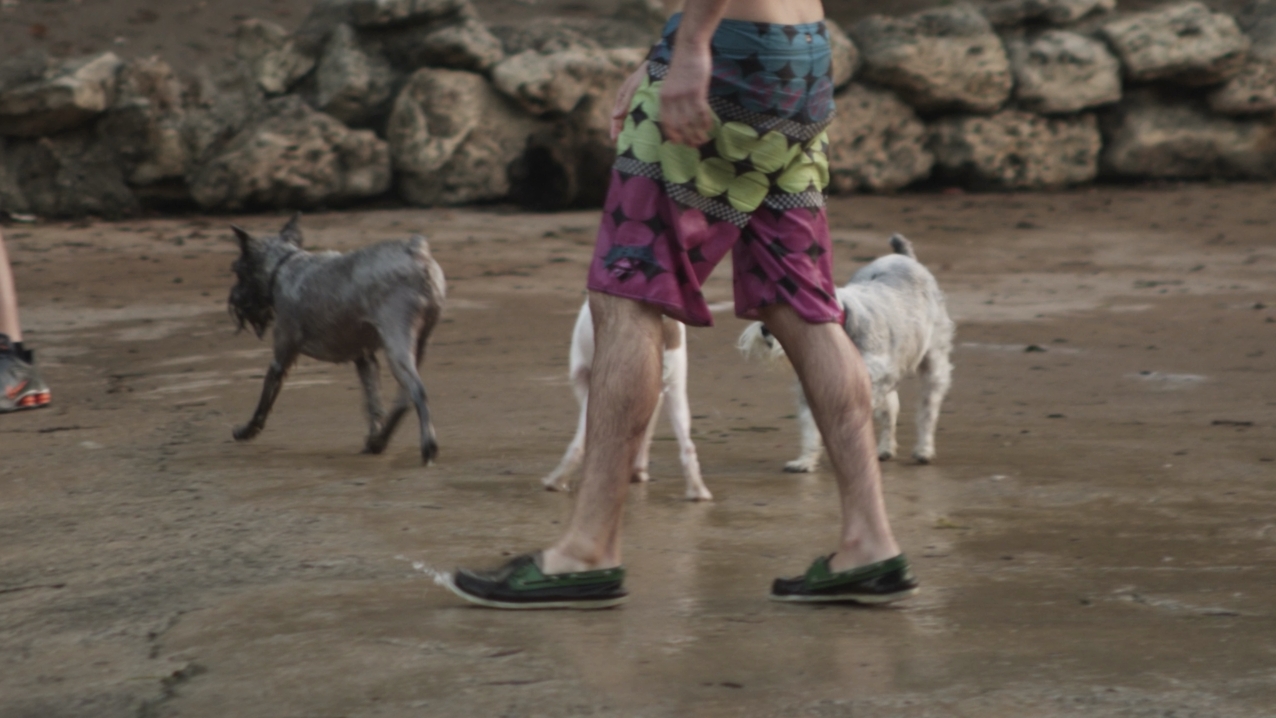}

}
\subfigure[si]{
\includegraphics[scale=0.132]{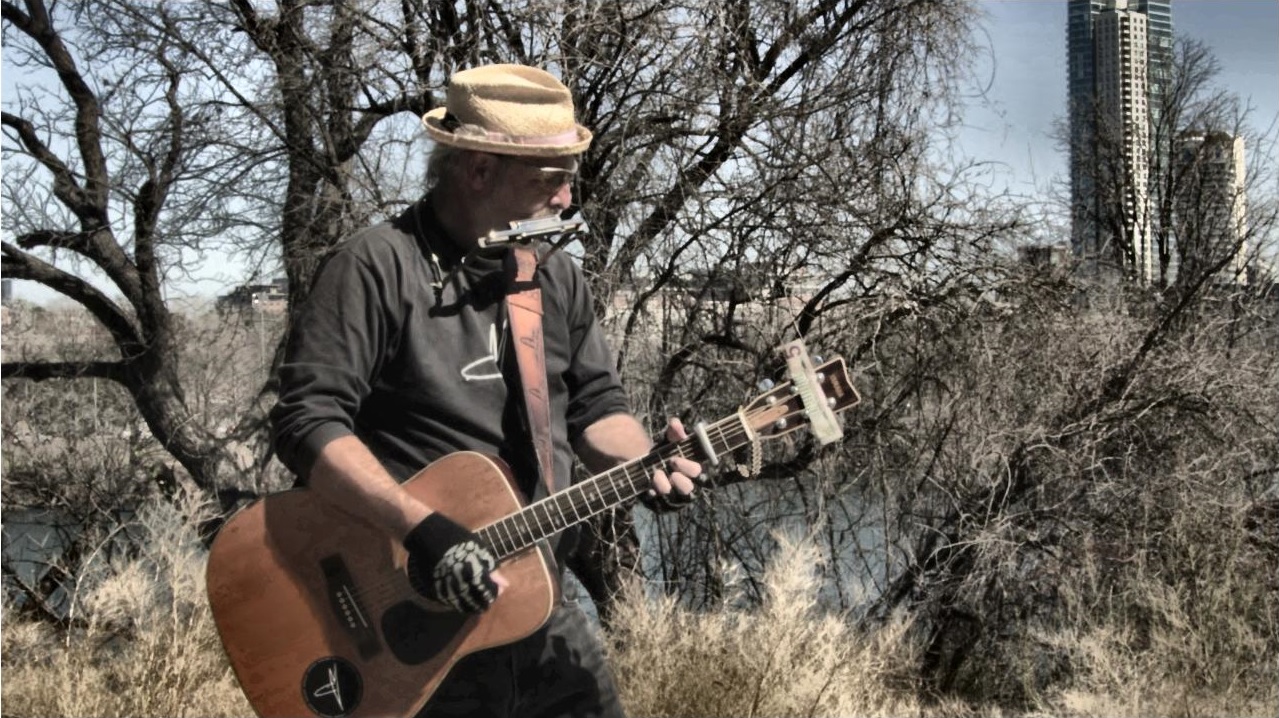}
}
\caption[Optional caption for list of figures]{
Sample frames of the video clips involved in the subjective study. The abbreviations of the names of the videos can be found in Table.~\ref{tab:video_descrpt}.
}
\label{fig:sampleframe}
\end{figure}

\item Using the video clips selected in the first step, 3 reference videos were constructed. They were used to generate quality-varying videos in the subjective study. Each reference video was constructed by concatenating 5 or 6 different clips (see Fig.~\ref{fig:refvideo}). The reference video were constructed in this way because long videos with monotonous content can be boring to subjects. This could adversely impact the accuracy of the TVSQ measured in the subjective study. The length of each video is 300 seconds, which was chosen to agree with the value recommended by the ITU \cite{ITU}. This is longer than the videos tested in \cite{Tan98,MasHem2004,SesBov11,NN,Anush10,KalpanaSPIE} and thus is a better tool towards understanding the long-term behaviorial responses of human vision system.
    \begin{figure}[h!]
    \centering
    \includegraphics[width=0.9\columnwidth]{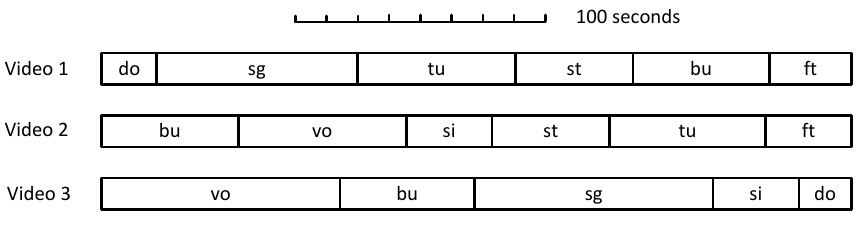}
    \caption{The construction of the reference videos. The abbreviation of the names of the clips can be found in Table.~\ref{tab:video_descrpt}. }
    \label{fig:refvideo}
    \end{figure}
\item For each reference video, 28 distorted versions were generated. Specifically, every reference video sequence was encoded into 28 constant bitrate streams using the H.264 encoder in \cite{ffmpeg} and then were decoded. To achieve a wide range of video quality exemplars, the encoding bitrates were chosen from hundreds of Kbps to several Mbps.
\item Every distorted version was partitioned into 1 second long video chunks and their STSQs were predicted with the computationally efficient and perceptually accurate RRED index \cite{RRED}. Let the RRED index of the $t^\mathrm{th}$ chunk in the $\ell^\mathrm{th}$ distorted version of the $k^\mathrm{th}$ reference video be denoted by $\mathrm{q}^\mathrm{rred}_{\ell,k}[t]$, where $t\in\{1,\dots,300\}$ second, $\ell\in\{1,\dots,28\}$, and $k\in\{1,2,3\}$. Then the Difference Mean Opinion Score (DMOS) of the STSQ for each chunk was predicted via logistic regression:
    \begin{equation}
    \label{eq:dmos}
    \mathrm{q}^\mathrm{dmos}_{\ell,k}[t]=16.4769+9.7111\log\left(1+\frac{\mathrm{q}^\mathrm{rred}_{\ell,k}[t]}{0.6444}\right).
    \end{equation}
    The regression model in \eqref{eq:dmos} was obtained by fitting a logistic mapping from the RRED index to the DMOSs on the LIVE Video Quality Assessment Database \cite{LIVEvideo}. Here, $\mathrm{q}^\mathrm{dmos}_{\ell,k}[t]$ ranges from 0 to 100 where lower values indicate better STSQ. To represent STSQ more naturally, so that higher numbers indicate better STSQ, a Reversed DMOS (RDMOS) is employed as follows:
    \begin{equation}
    \mathrm{q}^\mathrm{rdmos}_{\ell,k}[t] =  100-\mathrm{q}^\mathrm{dmos}_{\ell,k}[t].
    \end{equation}
    Broadly speaking, a RDMOS of less than 30 on the LIVE databases \cite{LIVEvideo} indicates bad quality, while scores higher than 70 indicate excellent quality. As an example, Fig.~\ref{fig:database}\subref{fig:videomake} plots $\mathrm{q}^\mathrm{rdmos}_{\ell,k}[t]$ for all of the distorted versions of the first reference video. Clearly, their STSQ covers a wide range of RDMOSs.
    \begin{figure}[h!]
    \centering
    \subfigure{\includegraphics[width=0.85\columnwidth]{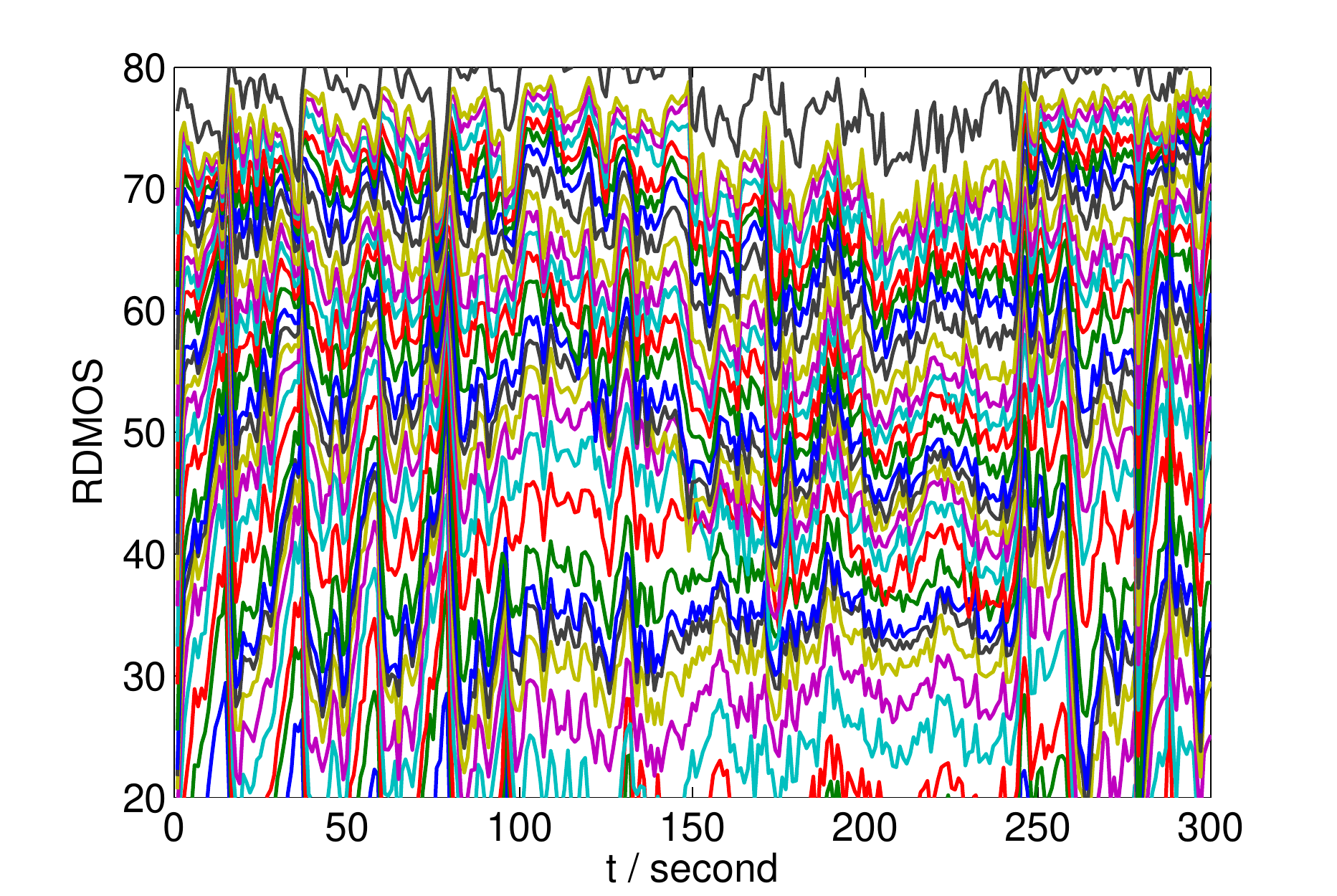}
    \label{fig:videomake}}
    \subfigure{\includegraphics[width=0.85\columnwidth]{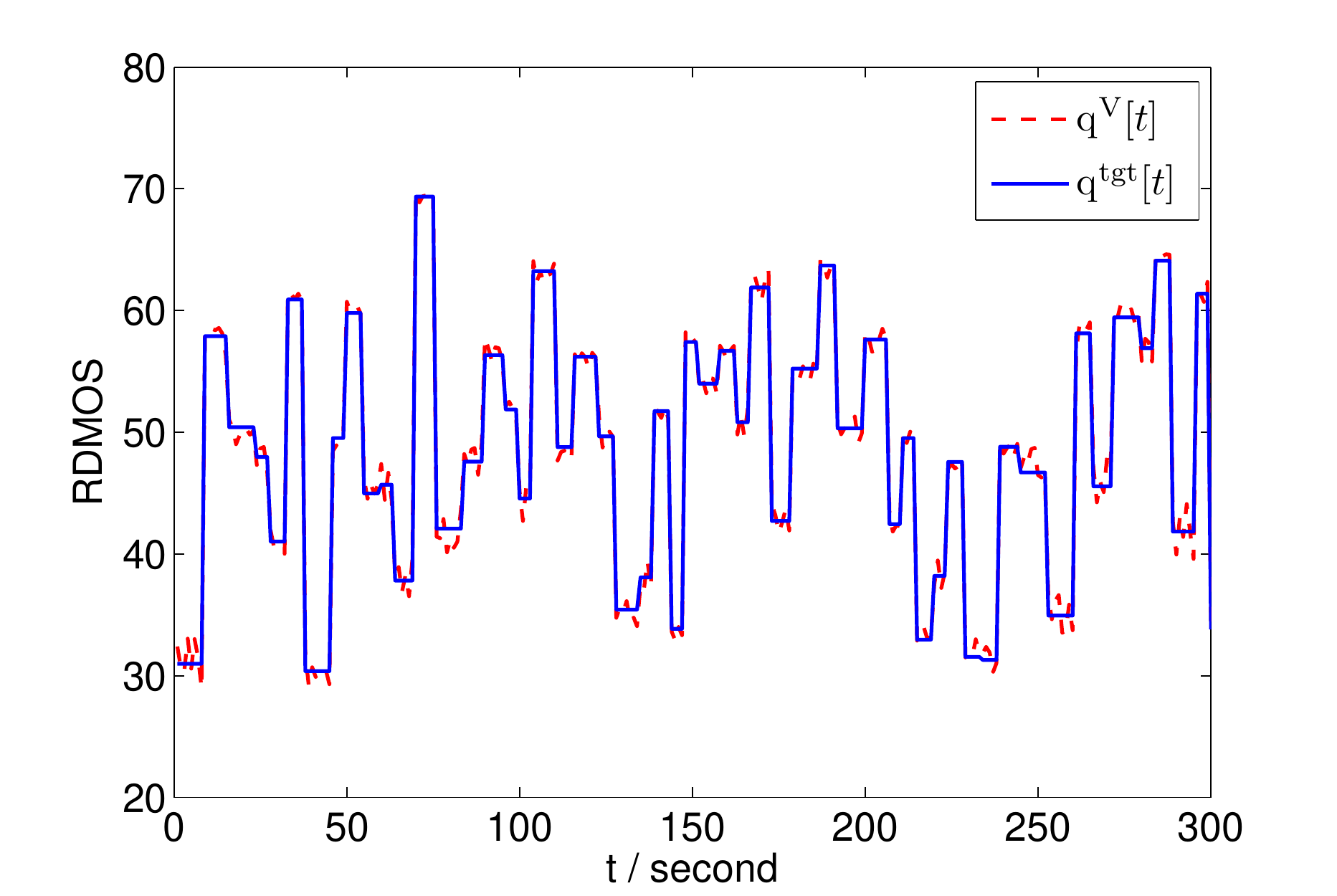}
    \label{fig:samplepath}}
    \caption{\subref{fig:videomake} The STSQ of each compressed version of the reference video is shown in different colors. \subref{fig:samplepath} A example of the designed target video quality $\mathrm{q^{tgt}}[t]$ and the actual video quality $\mathrm{q^{st}}[t]$ of the video sequence used in our database.}
    \label{fig:database}
    \end{figure}

\item Finally, for each reference video, 6 quality-varying videos were constructed by concatenating the video chunks selected from different distorted versions. For the $k^\mathrm{th}$ reference video, 6 target STSQ sequences $\left\{\big(\mathrm{q}^\mathrm{tgt}_{j,k}\big)_{1:300}, j=1,\dots,6\right\}$ were designed to simulate the typical quality variation patterns in HTTP-based streaming (see section \ref{sec:tgt} for more details). Then, 6 quality-varying videos were constructed such that their STSQs approximate the target sequences. Specifically, the $t^\mathrm{th}$ chunk of the $j^\mathrm{th}$ quality-varying video was constructed by copying the $t^\mathrm{th}$ chunk in the ${\ell^*_{t,j,k}}$-th distorted version, where
    \begin{equation}
        {\ell^*_{t,j,k}}=\arg\min_\ell\left|\mathrm{q}^\mathrm{tgt}_{j,k}[t]-\mathrm{q}^\mathrm{rdmos}_{\ell,k}[t]\right|.
    \end{equation}
    Denoting the STSQ of the $t^\mathrm{th}$ chunk in the obtained video by $\mathrm{q}^\mathrm{st}_{j,k}[t]$ , we have
    \begin{equation}
        \mathrm{q}^\mathrm{st}_{j,k}[t]=\mathrm{q}^\mathrm{rdmos}_{\ell^*_{t,j,k}}[t].
    \end{equation}
    As can be seen in Fig.~\ref{fig:database}, since the RDMOS scale is finely partitioned by the RDMOSs of the compressed versions, the error between the obtained STSQ $\mathrm{q}^\mathrm{st}_{j,k}[t]$ and the target STSQ $\mathrm{q}^\mathrm{tgt}_{j,k}[t]$ is small. Among the 6 quality-varying videos generated from each reference video, 1 video is used for subjective training and the other 5 videos are used for subjective test. In all, $3\times 1=3$ training videos and $3\times5=15$ test videos were constructed.
\end{list}

With this procedure, the pattern of quality variations in the test video sequences is determined by the target video quality sequence $\big(\mathrm{q}^\mathrm{tgt}_{j,k}\big)$. The design of $\big(\mathrm{q}^\mathrm{tgt}_{j,k}\big)$ is described next.


\subsection{Target Video Quality Design}
\label{sec:tgt}
\begin{figure}[h!]
\centering
\includegraphics[width=0.6\columnwidth]{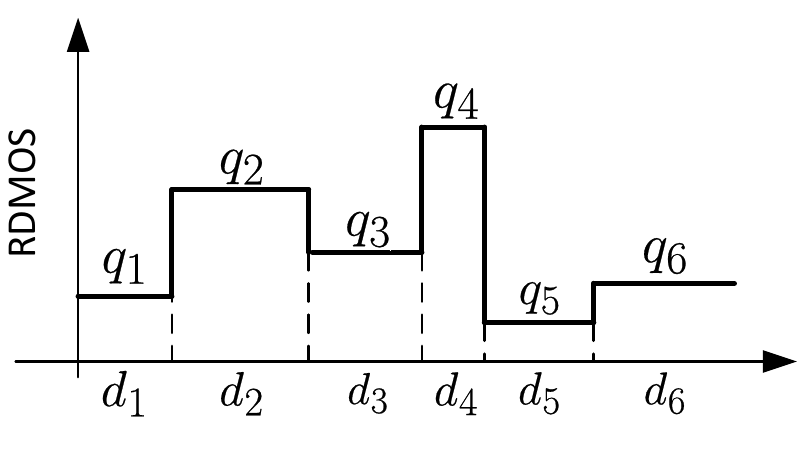}
\caption{The design of the target STSQs. The durations of each segment $d_1, d_2, \dots$ were realizations of $\mathsf{D}(1),\mathsf{D}(2), \dots$. The STSQ levels $q_1, q_2, \dots$ were realizations of $\mathsf{Q}(1),\mathsf{Q}(2), \dots$.}
\label{fig:target}\end{figure}

To obtain a good TVSQ prediction model for videos streamed over HTTP, the target video quality $\big(\mathrm{q}^\mathrm{tgt}_{j,k}\big)_{1:300}$ was designed such that the generated quality-varying videos can roughly simulate the STSQs of videos streamed over HTTP. In HTTP-based video streaming protocols such as those described in   \cite{smoothstream,livestream,dynamicsstream,DASH}, videos are encoded into multiple representations at different bitrates. Each representation is then partitioned into segments, each several seconds long. The client dynamically selects a segment of a representation to download. Therefore, in our subjective study, $\big(\mathrm{q}^\mathrm{tgt}_{j,k}\big)_{1:300}$ was designed as a piece-wise constant time-series. Specifically, $\big(\mathrm{q}^\mathrm{tgt}_{j,k}\big)_{1:300}$ was generated using two independent random processes. The first random process $\{\mathsf{D}(s):s=1,2,\dots\}$ simulates the length of the video segments. The second random process $\{\mathsf{Q}(s): s=1,2,\dots\}$ simulates the STSQs of segments. The sequence $\big(\mathrm{q}^\mathrm{tgt}_{j,k}\big)_{1:300}$ was constructed as a series of constant-value segments where the durations of the segments were given by $\mathsf{D}(s)$ and the RDMOSs of the segments were given by $\mathsf{Q}(s)$ (see Fig.~\ref{fig:target}).

In HTTP-based video streaming protocols, the duration of video segments can be flexibly chosen by the service provider. Shorter durations allow more flexibility for rate adaptation when the channel condition is varying rapidly. For example, due to the mobility of wireless users, the wireless channel throughput may vary on time scales of several seconds \cite{Goldsmith}. Consequently, this work focus on applications where the lengths of the segments are less than 10 seconds. TVSQ modeling for videos undergoing slowly varying data rates has been investigated in \cite{MasHem2004}\cite{NN}\cite{Tan98}. In a subjective experiment, there is always a delay or latency between a change in STSQ and a subject's response. During the experimental design, we found that if the video quality varied too quickly, subjects could not reliably track their judgments of quality to the viewed videos. Specifically, when the video quality changes, a subject may take 1-2 seconds to adjust his/her opinion on the TVSQ. If the quality is adapted frequently, the quality variations that occur during this adjustment process can annoy the subject and thus reduce the accuracy of the measured TVSQs. Thus, we restricted the length of each segment to be at least 4 seconds, which is comfortably longer than the subjects' latency and short enough to model quality variations in adaptive video streaming. In sum, the random process $\{\mathsf{D}(s):s=1,2,\dots\}$ takes values from the set $\{4,5,6,7,8,9,10\}$.

The distribution of STSQs of a video transported over HTTP depends on many factors including the encoding bitrates, the rate-quality characteristics, the segmentation of each representation, the channel dynamics, and the rate adaptation strategy of the client. To sample uniformly from among all possible patterns of STSQ variations, the random processes $\mathsf{D}(s)$ and $\mathsf{Q}(s)$ were designed as i.i.d. processes, which tend to traverse all possible patterns of quality variations. Also, the distributions of $\mathsf{D}(s)$ and $\mathsf{Q}(s)$ were designed to ``uniformly" sample all possible segment lengths and STSQ levels, respectively. To this end, we let $\mathsf{D}(s)$ take values in the set $\{4,5,6,7,8,9,10\}$ with equal probability. Similarly, the distribution of $\mathsf{Q}(s)$ was designed such that the sample values of $\mathsf{Q}(s)$ would be distributed as if the videos were uniformly sampled in the LIVE database, because that set of videos is carefully chosen to represent a wide range of perceptually separated STSQ \cite{LIVEvideo}. The RDMOSs of videos in the LIVE database are distributed as approximately obeying a normal distribution $\mathcal N(50,10^2)$ \cite{LIVEvideo}. Therefore, we let the distribution of $\mathsf{Q}(s)$ be $\mathcal N(50,10^2)$. In the LIVE database, almost all of the recorded RDMOSs fall within the range $[30,70]$. Videos with RDMOS lower than 30 are all very severely distorted while videos with RDMOS higher than 70 are all of high quality. Due to saturation of the subjects' scoring capability outside these ranges, the recorded qualities of videos with RDMOSs lower than 30 or higher than 70 are not separable. Therefore, we truncated $\mathsf{Q}(s)$ to the range $[30,70]$.




\subsection{Subjective Experiments}
\label{sec:experiment}
A subjective study was conducted to measure the TVSQs of the quality-varying videos in our database. The study was completed at the LIVE subjective testing lab at The University of Texas at Austin. The videos in our database were grouped into 3 sessions. Each session included one of the three reference videos and the 6 quality-varying videos generated from the reference video. The videos in each session were each viewed and scored by 25 subjects. One of the quality-varying videos was used as a training sequence. The other six videos, including 5 quality-varying videos and the reference video, were used for subjective study.

\begin{figure}[h!]
\centering
\includegraphics[width=\columnwidth]{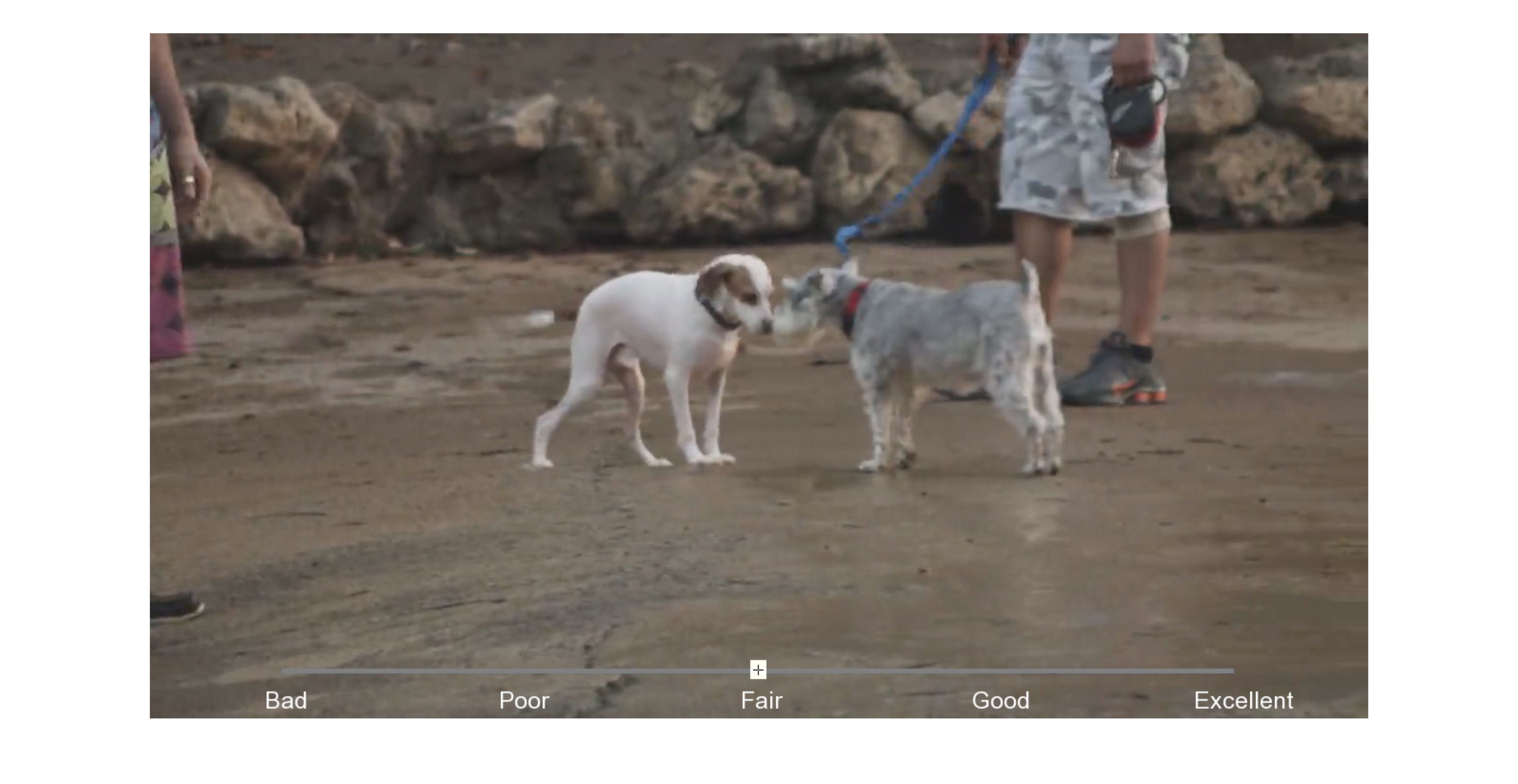}
\caption{User interface used in the subjective study.}
\label{fig:interface}
\end{figure}

A user interface was developed for the subjective study using the Matlab XGL toolbox \cite{XGL}. The user interface ran on a Windows PC with an Intel Xeon 2.93GHz CPU and a 24GB RAM. The XGL toolbox interfaced with ATI Radeon X300 graphics card on the PC to precisely display video frames without latencies or frame drops, by loading each video into memory before display. Video sequences were displayed to the viewers on a 58 inch Panasonic HDTV plasma monitor at a viewing distance of about 4 times the picture height. During the play of each video, a continuous scale sliding bar was displayed near the bottom of the screen. Similar to the ITU-R ACR scale \cite{ITU}, the sliding bar was marked with five labels: ``Bad", ``Poor", ``Fair", ``Good", and ``Excellent", equally spaced from left to right. The subject could continuously move the bar via a mouse to express his/her judgment of the video quality as each video is played. The position of the bar was sampled and recorded automatically in real time as each frame is displayed (30 fps). No mouse clicking was required in the study. Fig.~\ref{fig:interface} shows the subjective study interface including a frame of a displayed video.

During the training period, each subject first read instructions describing the operation of the user interface (see Appendix \ref{app:instructions}), then practiced on the training sequence. The subject then started rating the test videos (reference video and five quality-varying videos) shown in random order. The subjects were unaware of the presence of the reference video.
\subsection{Data Preprocessing}
Denote the average score assigned by the $i^\mathrm{{th}}$ subject to the frames of the $t^\mathrm{th}$ chunk of the $j^\mathrm{th}$ quality-varying video in the ${k}^\mathrm{th}$ session by $\mathrm{c}_{i,j,k}[t]$. Let the score assigned to the reference video be denoted by $\mathrm{c}^\mathrm{ref}_{i,k}[t]$. The impact of video content was offseted on the TVSQs of the test videos using
\begin{equation}
\mathrm{c}_{i,j,k}^\mathrm{offset}[t]=100-\left(\mathrm{c}^\mathrm{ref}_{i,k}[t]-\mathrm{c}_{i,j,k}[t]\right).
\end{equation}
Here, $\mathrm T$ denotes the length of the test videos and $\mathrm J$ denotes the number of quality-varying videos in each session. In our experiment, ${\mathrm T}=300$ and $\mathrm J=5$. Note that the subjects deliver their quality judgments in real-time as the test video is being displayed. To avoid distracting the subjects from viewing the video, we did not require them to use the full scale of the sliding bar. Moreover, such an instruction may tend to bias the recorded judgments from their natural response. Thus, each subject was allowed to freely deploy the sliding bar when expressing their judgments of TVSQ. To align the behavior of different subjects, paralleling to prior work such as \cite{Liu10,Ribeiro11,Lin12,wirelessVQA,Sheikh06}, we normalize $\big(\mathrm{c}_{i,j,k}^\mathrm{offset}\big)$ by computing the Z-scores \cite{Zscore} as follows:
\begin{equation}
\label{eq:zscore}
\begin{aligned}
  m_{i,k}&=\frac{1}{\mathrm J}\frac{1}{\mathrm T}\sum_{j=1}^{\mathrm J}\sum_{t=1}^{\mathrm T} \mathrm c_{i,j,k}^\mathrm{offset}[t]; \\
  \sigma^2_{i,k}&=\frac{1}{\mathrm{JT}-1}\sum_\mathrm{j=1}^{\mathrm J}\sum_{t=1}^{\mathrm T} \left(\mathrm{c}^\mathrm{offset}_{i,j,k}[t]-m_{i,k}\right)^2;\\
  \mathrm{z}_{i,j,k}[t]&=\frac{\mathrm{c}^\mathrm{offset}_{i,j,k}[t]-m_{i,k}}{\sigma_{i,k}}.
\end{aligned}
\end{equation}
In \eqref{eq:zscore}, the values of $m_{i,k}$ and $\sigma_{i,k}$ are respectively the mean and the variance of the scores assigned by the $i^\mathrm{th}$ subject in the $k^\mathrm{th}$ session. The value of $\mathrm{z}_{i,j,k}[t]$ is the normalized score. Let $\mathrm I$ denote the number of subjects. We have $\mathrm{I}=25$. Then for the $t^\mathrm{th}$ second of the ${j^\mathrm{th}}$ test video, the average and standard deviation of the Z-scores assigned by the subjects were computed
\begin{equation}
\begin{aligned}
  \mu_{j,k}[t]&=\frac{1}{\mathrm I}\sum_{i=1}^{\mathrm I} \mathrm{z}_{i,j,k}[t];\\
  \eta^2_{j,k}[t]&=\frac{1}{{\mathrm I}-1}\sum_{i=1}^{\mathrm I} \left(\mathrm{z}_{i,j,k}[t]-\mu_{j,k}[t]\right)^2.
\end{aligned}
\end{equation}
If $\mathrm{z}_{i,j,k}[t]>\mu_{j,k}[t]+2\eta_{j,k}[t]$ or $\mathrm{z}_{i,j,k}[t]<\mu_{j,k}[t]-2\eta_{j,k}[t]$, $\mathrm{z}_{i,j,k}[t]$ was marked as an outlier because the Z-score given by subject $i$ deviates far from the Z-scores given by the other subjects. The outliers were excluded and the Z-scores were recomputed using \eqref{eq:zscore}. Let $\mathcal O_{j,k,t}$ denote the set of subjects who assigned outlier Z-scores to the $t^\mathrm{th}$ chunk of the $j^\mathrm{th}$ video in the $k^\mathrm{th}$ session. The averaged Z-score of the TVSQ for the $t^\mathrm{th}$ chunk is then
\begin{equation}
\bar{\mathrm{z}}_{j,k}[t]=\frac{1}{\mathrm{I}-|\mathcal O_{j,k,t}|}\sum_{i\notin \mathcal O_{j,k,t}} \mathrm{z}_{i,j,k}[t].
\end{equation}
The $95\%$ confidence interval of the average Z-scores is $\bar{\mathrm{z}}_{j,k}[t]\pm 1.96\eta_{j,k}[t]/\sqrt{\mathrm{I}-|\mathcal O_{j,k,t}|}$. We found that the values of the averaged Z-scores all lie in the range $[-4,4]$. Therefore, $\bar{\mathrm{z}}_{j,k}[t]$ was mapped to the range $[0,100]$ using the following formula:
\begin{equation}
\mathrm{q}^\mathrm{tv}_{j,k}[t]=\frac{\bar{\mathrm{z}}_{j,k}[t]+4}{8}\times100.
\end{equation}
Correspondingly, the $95\%$ confidence interval of TVSQ is $\mathrm{q}^\mathrm{tv}_{j,k}[t]\pm\epsilon_{j,k}[t]$,
where
\begin{equation}
\label{eq:ci}
\epsilon_{j,k}[t]=\frac{1.96\eta_{j,k}[t]/\sqrt{\mathrm{I}-|\mathcal O_{j,k,t}|}+4}{8}\times100.
\end{equation}
In all, the TVSQ for $\mathrm N=3\times5=15$ quality-varying videos were measured. In the following, we replace the subscript $({j,k})$ with a subscript $1\leq{n}\leq \mathrm N$ to index the quality-varying videos and denote by $\mathrm{q}^\mathrm{tv}_n[t]$ and $\epsilon_{n}[t]$ the measured TVSQ and the confidence interval of the ${n^\mathrm{th}}$ video, respectively. Similarly, the STSQ of the ${n^\mathrm{th}}$ video predicted by the Video-RRED algorithm \cite{RRED} is denoted by $\mathrm{q}^\mathrm{st}_n[t]$.
\subsection{Preliminary Observations}
\label{sec:pre_obs}
\begin{figure}[h!]
\centering
\includegraphics[width=0.9\columnwidth]{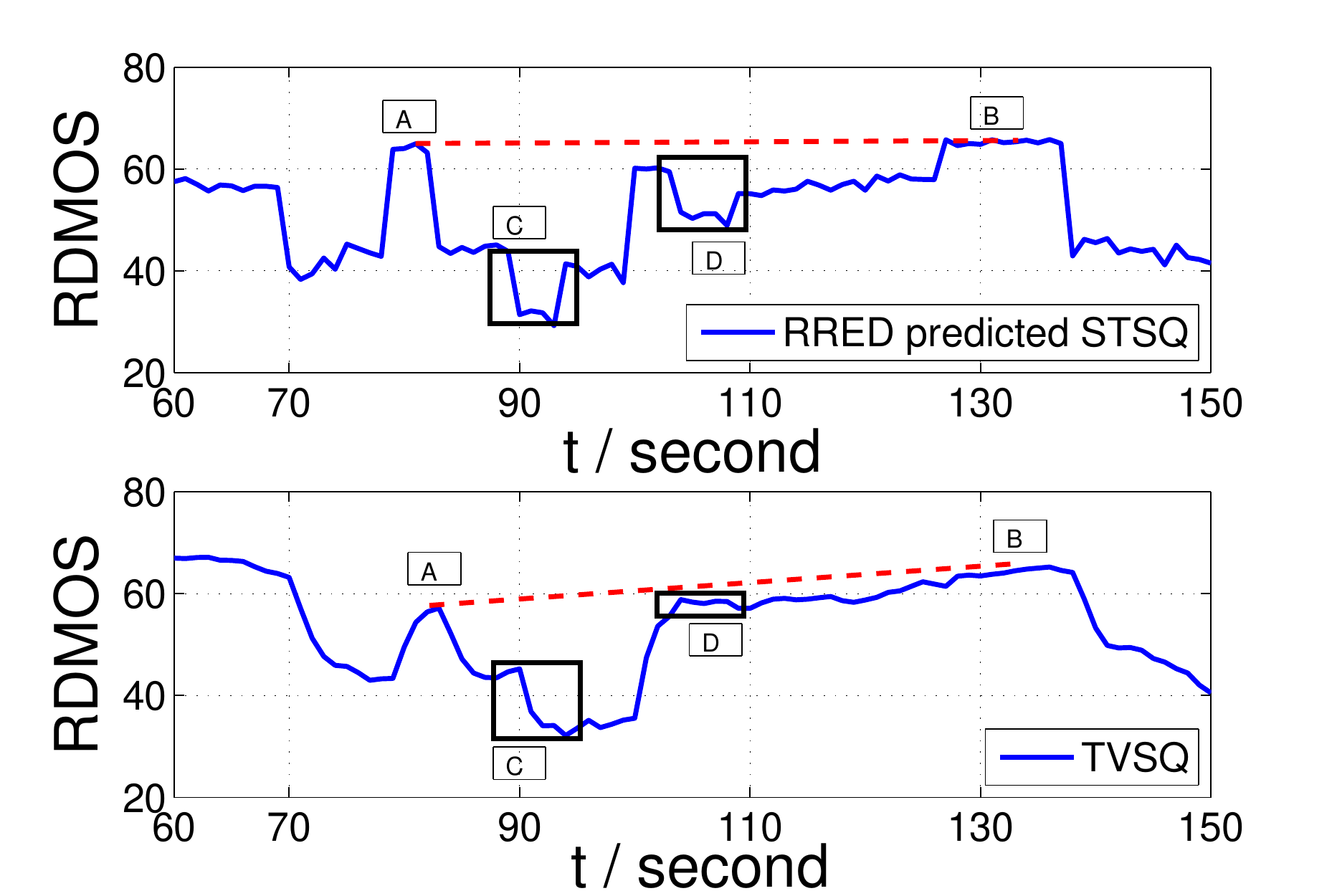}
\caption{(Upper) the STSQs predicted using Video-RRED. (Lower) The TVSQs measured in the subjective study.}\label{fig:pre_obs}
\end{figure}

Since $\left(\mathrm{q}^\mathrm{st}\right)$ is the predicted STSQ, we expect $\left(\mathrm{q}^\mathrm{st}\right)$ to contain useful evidence about the TVSQ. The $\left(\mathrm{q}^\mathrm{st}\right)$ and the corresponding $\left(\mathrm{q}^\mathrm{tv}\right)$ of the $6^\mathrm{th}$ quality-varying video from $t=61$ to $t=150$ is plotted in Fig.~\ref{fig:pre_obs}. It is seen that both the $\left(\mathrm{q}^\mathrm{st}\right)$ and the $\left(\mathrm{q}^\mathrm{tv}\right)$ follow the similar trend of variation. But it should be noted that the relationship between $\left(\mathrm{q}^\mathrm{st}\right)$ and $\left(\mathrm{q}^\mathrm{tv}\right)$ cannot be simply described by a static mapping. For example, at point A ($t=29$) and point B ($t=85$), the $\mathrm{q}^\mathrm{st}[t]$ takes similar values. But the corresponding $\mathrm{q}^\mathrm{tv}[t]$ is lower at point A than point B. This observation could be explained by the hysteresis effects. Prior to point A, $\mathrm{q}^\mathrm{st}[t]$ is around 40 (see $\left(\mathrm{q}^\mathrm{st}\right)_{20:28}$). But, prior to point B, $\mathrm{q}^\mathrm{st}[t]$ is around 65 (see $\left(\mathrm{q}^\mathrm{st}\right)_{76:84}$). Thus, the previous viewing experience is worse at point A, which gives rise to a lower TVSQ. Such hysteresis effects should be considered in HTTP-based rate adaptations. For example, if the ``previous viewing experience" is bad (such as point A), the server should send the video segment of higher quality to counterbalance the impact of bad viewing experience on the TVSQ.

It may be observed that the $\mathrm{q}^\mathrm{st}[t]$ experiences the similar level of drop in region C and region D. The drop of $\mathrm{q}^\mathrm{st}[t]$ in region C results in a significant drop in $\mathrm{q}^\mathrm{tv}[t]$. But, in region D, $\mathrm{q}^\mathrm{st}[t]$ is not as affected by the drop of $\mathrm{q}^\mathrm{st}[t]$. In other words, the sensitivity of TVSQ to the variation in $\left(\mathrm{q}^\mathrm{st}\right)$ is different in region C and region D. This is probably due to the non-linearities of human behavioral responses. Including such nonlinearities is critical for efficient HTTP-based adaptation. Specifically, when the TVSQ is insensitive to the STSQ (such as in region D), the server may switch to a lower streaming bitrate to reserve some resources (such as transmission time) without hurting the TVSQ. Those reserved resources can then be used to maintain a good TVSQ when the TVSQ is sensitive (such as region C).

In sum, quantitatively modeling the hysteresis effects and the nonlinearities are critical for TVSQ-optimized rate adaptations. This motivate us to propose a non-linear dynamic system model, which is described in more detail below.

\section{System Model Identification}
\label{sec:model_id}
In this section, the model for TVSQ prediction is presented in Section~\ref{sec:Modelintro}. Then, the algorithm for model parameter estimation is described in Section~\ref{sec:model_idalgo}. The method for model order selection is introduced in Section~\ref{sec:model_order}.
\subsection{Proposed Model for TVSQ Prediction}
\label{sec:Modelintro}

\begin{figure}[h!]
\centering
\includegraphics[width=12cm]{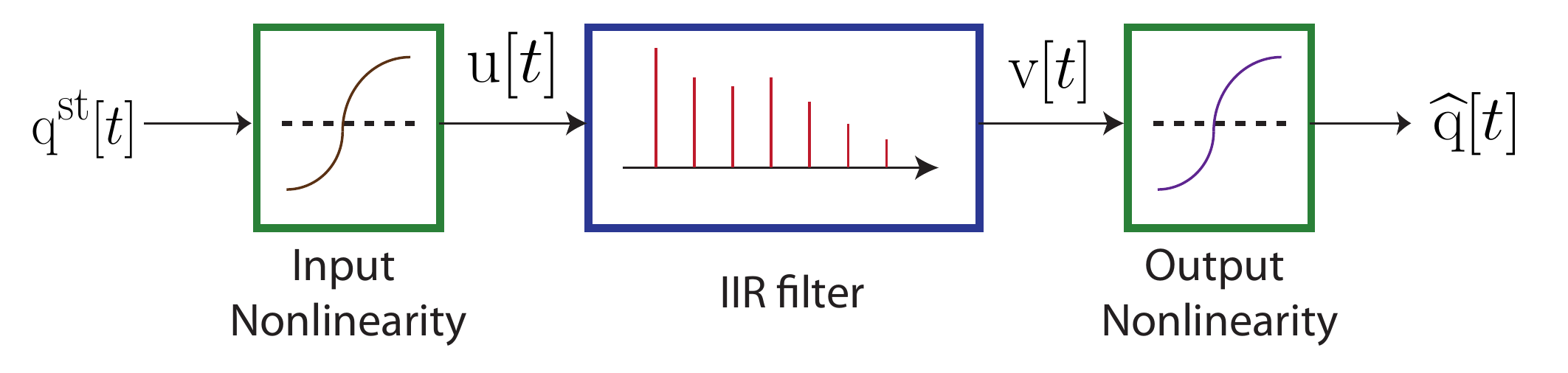}
\caption{Proposed Hammerstein-Wiener model for TVSQ prediction.}
\label{fig:hwmodel}
\end{figure}

Due to the hysteresis effect of human behaviorial responses to quality variations, the TVSQ at a moment depends on the viewing experience prior to the current moment. A dynamic system model can be used to capture the hysteresis effect using the ``memory" of the system state. The simplest type of dynamic system is a linear filter. The human vision system, however, is non-linear in general \cite{Watson97}\cite{Heeger94}\cite{Daly93}. Although introducing intrinsic nonlinearities into the dynamic system model could help to capture those nonlinearities\footnote{We say a nonlinear system has intrinsic nonlinearity if its current system state is a nonlinear function of the previous system state and input. Otherwise, we say the system has extrinsic nonlinearity.}, the dynamic system would become too complicated to provide guidance on the design of TVSQ-optimized rate-adaptation algorithms. More specifically, due to the randomness of channel conditions, the TVSQ-optimized rate-adaptation algorithm design is essentially a stochastic optimization problem. For a linear dynamic model with input $\big(\mathrm{x}\big)$, its output $\big(\mathrm{y}\big)$ is given by $\big(\mathrm{y}\big)=\big(\mathrm{h}\big)*\big(\mathrm{x}\big)$, where $\big(\mathrm{h}\big)$ is the impulse response. Due to the linearity of expectation, the expectation of the TVSQs can be characterized using $\mathbb{E}[y]=||\mathbf{h}||_1\mathbb{E}[x]$. For a dynamic model with intrinsic nonlinearities, however, linearity of expectation cannot be applied and analyzing the average behavior of the TVSQ becomes difficult. Therefore, we employed a Hammerstein-Wiener (HW) model \cite{Sysid}, which captures the nonlinearity with extrinsic nonlinear functions. The model is illustrated in Fig.~\ref{fig:hwmodel}. The core of the HW model is a linear filter (see \cite{Sysid}) which is intended to capture the hysteresis. At the input and output of the HW model, two non-linear static functions are employed to model potential non-linearities in the human response. We call these two functions input nonlinearity and output nonlinearity, respectively.

The linear filter has the following form:
\begin{equation}
\label{eq:oe}
\begin{aligned}
\mathrm{v}[t]&=\sum_{d=0}^r b_d\,\mathrm{u}[t-d]+\sum_{d=1}^r f_d\,\mathrm{v}[t-d]\\
    &={\bf b}^\mathsf{T}\big(\mathrm{u}\big)_{t-r:t}+{\bf f}^\mathsf{T}\big(\mathrm{v}\big)_{t-r:t-1},
\end{aligned}
\end{equation}
where the parameter $r$ is the model order and the coefficients ${\bf b}=(b_0,\dots,b_r)^\mathsf{T}$ and ${\bf f}=(f_1,\dots,f_r)^\mathsf{T}$ are model parameters to be determined. At any time $t$, the model output $\mathrm{v}[t]$ depends not only on the previous $r$ seconds of the input $\mathrm{u}[t]$, but also on the previous $r$ seconds of $\mathrm{v}[t]$ itself. Thus this filter has an infinite impulse response (IIR). We employed this model rather than a finite impulse response (FIR) filter because the IIR filter can model the long-term impact of quality variations with a lower model order and thus using fewer parameters. To train a parameterized model, the size of the training data set increases exponentially with the number of the parameters \cite{Sysid}. Therefore, it is easier to train an IIR model. A drawback of the IIR filter \eqref{eq:oe} is its dependency on its initial state. Specifically, to compute $\big(\mathrm{v}\big)_{t>r}$, the initial $r$ seconds of output $\big(\mathrm{v}\big)_{1:r}$ need to be known. But $\big(\mathrm{v}\big)_{1:r}$ is the TVSQ of the user, which is unavailable. Actually, it can be shown that this unknown initial condition only has negligible impact on the performance of the proposed model. A more detailed analysis is presented in Section~\ref{sec:ini_state}.

To model the input and output nonlinearities of the HW model, we have found that if the input and output static functions are chosen as generalized sigmoid functions \cite{generalized_log}, then the proposed HW model can predict TVSQ accurately. Thus, the input and output functions were set to be
\begin{align}
\label{eq:input}
\mathrm{u}[t]=\beta_3+\beta_4\frac{1}{1+\exp\left(-(\beta_1\mathrm{q^{st}}[t]+\beta_2)\right)},
\end{align}
and
\begin{align}
\label{eq:output}
\widehat{\mathrm{q}}[t]=\gamma_3+\gamma_4\frac{1}{1+\exp\left(-(\gamma_1\mathrm{v}[t]+\gamma_2)\right)},
\end{align}
where ${\boldsymbol\beta}=(\beta_1,\dots,\beta_4)^\mathsf{T}$ and ${\boldsymbol\gamma}=(\gamma_1,\dots,\gamma_4)^\mathsf{T}$ are model parameters and $\widehat{\mathrm{q}}$ is the predicted TVSQ. 

Let ${\boldsymbol\theta}=({\bf b}^\mathsf{T}, {\bf f}^\mathsf{T}, {\boldsymbol\beta}^\mathsf{T}, {\boldsymbol\gamma}^\mathsf{T})^\mathsf{T}$ be the parameters of the proposed HW model, and let $\widehat{\mathrm{q}}$ be regarded as a function both of time $t$ and parameter $\boldsymbol\theta$. Thus, in the following, we explicitly rewrite $\widehat{\mathrm{q}}$ as $\widehat{\mathrm{q}}(t,\boldsymbol\theta)$. To find the optimal HW model for TVSQ prediction, two things need to be determined: the model order $r$ and the model parameter $\boldsymbol\theta$. In the following, we first show how to optimize the model parameter $\boldsymbol\theta$ for given model orders. Then, we introduce the method for model order estimation.

\subsection{Model Parameter Training}
\label{sec:model_idalgo}

This section discusses how to optimize the model parameter $\boldsymbol\theta$
such that the error between the measured TVSQ and the predicted TVSQ can be minimized. In system identification and machine learning, the mean square error (MSE) is the most widely used error metric. Denoting the predicted TVSQ of the $n^\mathrm{{th}}$ video by $\widehat{\mathrm{q}}_n(t,\boldsymbol\theta)$, the MSE is defined as $\frac{1}{\mathrm{N}\mathrm{T}}\sum_{{n}=1}^\mathrm{N}\sum_{t=1}^{\mathrm T}\left(\widehat{\mathrm{q}}_{n}(t,\boldsymbol\theta)-\mathrm{q}_{n}^\mathrm{tv}[t]\right)^2$. The MSE always assigns a higher penalty to a larger estimation error. For the purposes of tracking TVSQ, however, once the estimated TVSQ deviates far from the measured TVSQ, the model fails. There is no need to penalize a large error more than another smaller, yet still large error. Furthermore, since the $\mathrm{q}^\mathrm{tv}_n[t]$ is just the average subjective quality judgment, the confidence interval of TVSQ $\epsilon_n[t]$ (see the definition in \eqref{eq:ci}) should also be embodied in the error metric to account for the magnitude of the estimation error. We chose to use the outage rate, also used in \cite{NN}, as the error metric. Specifically, the outage rate of a TVSQ model is defined as the frequency that the estimated TVSQ deviates by at least twice the confidence interval of the measured TVSQ. More explicitly, outage rate can be written as

\begin{align}
\label{eq:or}
\mathrm{E}(\boldsymbol\theta)=\frac{1}{\mathrm{N}\mathrm{T}}\sum_{{n}=1}^\mathrm{N}\sum_{t=1}^{\mathrm T}{\mathbbm 1}{\left(\left|\widehat{\mathrm{q}}_n(t,\boldsymbol\theta)-\mathrm{q}^\mathrm{tv}_n[t]\right|>2\epsilon_n[t]\right)},
\end{align}
where $\mathbbm 1(\cdot)$ is the indicator function.

\begin{figure}[!t]
\centering
\includegraphics[width=0.75\columnwidth]{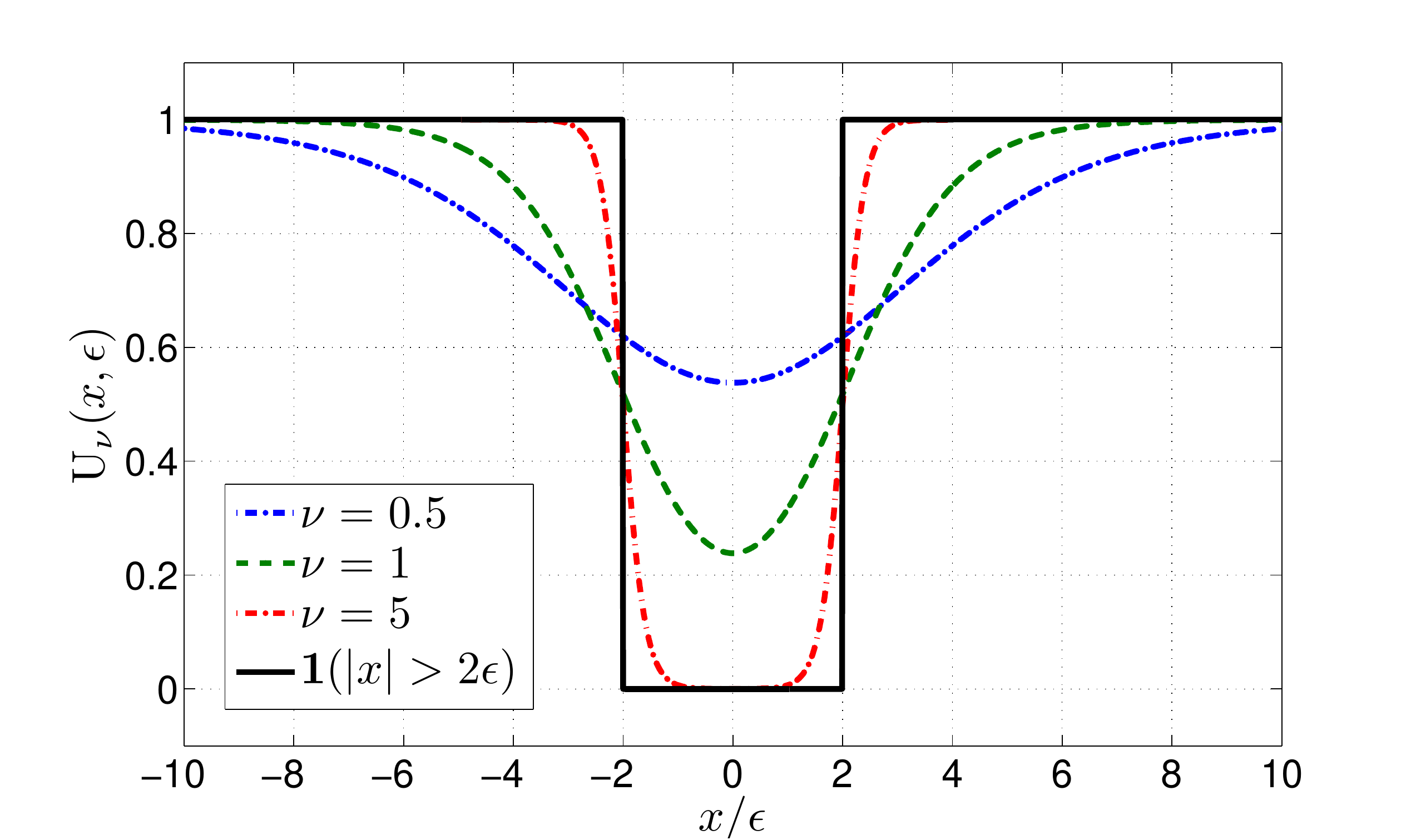}
\caption[Optional caption for list of figures]{
${\mathrm{ U}_{\nu}(x,\epsilon)}$ with $\nu=0.5$, 1 and 5.
}\label{fig:E_u}
\end{figure}

Gradient-descent parameter search algorithms are commonly used for model parametrization. In our case, however, the gradient of the indicator function $\mathbbm 1(\cdot)$ in \eqref{eq:or} is zero almost everywhere and thus the gradient algorithm cannot be applied directly. To address this difficulty, we approximated the indicator function $\mathbbm 1(|x|>2\epsilon)$ by a penalty function
\begin{align}
\label{eq:u_apprx}
{\mathrm{ U}_{\nu}(x,\epsilon)}
=\mathrm{h}(x,\nu,-2\epsilon)+\left(1-\mathrm{h}(x,\nu,2\epsilon)\right),
\end{align}
where $\mathrm{h}(x,\alpha,\zeta)=1/\left(1+\exp(-\alpha(x+\zeta)\right)$ is a logistic function. In Fig.~\ref{fig:E_u}, ${\mathrm{ U}_{\nu}(x,\epsilon)}$ with different configurations of the parameter $\nu$ are plotted. It can be seen that, as $\nu\rightarrow\infty$, ${\mathrm{ U}_{\nu}(x,\epsilon)}$ converges to $\mathbbm1\left(|x|>2\epsilon\right)$.
The outage rate $\mathrm{E}(\boldsymbol\theta)$ can thus be approximated by $\mathrm{E}(\boldsymbol\theta)=\lim_{\nu\rightarrow\infty}\mathrm{E^{apx}_{\nu}}(\boldsymbol\theta)$, where
\begin{align}
\mathrm{E^{apx}_{\nu}}(\boldsymbol\theta)=\frac{1}{\mathrm{N}\mathrm{T}}\sum_{{n}=1}^\mathrm{N}\sum_{t=1}^{\mathrm T}{\mathrm {U}_{\nu}\left(\widehat{\mathrm{q}}_n(t,\boldsymbol\theta)-\mathrm{q}^\mathrm{tv}_n[t],\epsilon[t]\right)}.
\end{align}
The iterative algorithm used for model parameter identification is described in Algorithm~\ref{alg:sysid}. In the $i^\mathrm{th}$ iteration, a gradient-descent search algorithm is applied to minimize $\mathrm{E^{apx}_{\nu}}(\boldsymbol\theta)$. The obtained parameter $\boldsymbol\theta^i$ is then used as the starting point for the gradient-descent search in the $(i+1)^\mathrm{th}$ iteration. At the end of each iteration, the parameter $\nu$ is increased by $\nu:=1.2\nu$ \footnote{The choice of the multiplicative factor 1.2 is to balance the efficiency and accuracy of the algorithm. Any number less than 1.2 gives rise to similar performance. Any number larger than 1.2 results in worse performance.}. Using this algorithm, the penalty function $\mathrm{U}_\nu(x,\epsilon)$ is gradually modified to $\mathbbm 1(|x|>2\epsilon)$ such that the estimated TVSQ is forced into the confidence interval of the measured TVSQ. Note that when $\nu\geq20$, ${\mathrm{U}_{\nu}(x,\epsilon)}$ is very close to $\mathbbm1\left(|x|>2\epsilon\right)$. Hence, the iteration is terminated when $\nu\geq20$ \footnote{Since $\mathrm{E}(\boldsymbol\theta)$ is not a convex function of $\theta$, gradient-descent can only guarantee local optimality.}.
\begin{algorithm}[!t]
\caption{Parameter optimization algorithm}
\label{alg:sysid}
\begin{algorithmic}[1]
\Require $\mathrm{q}^\mathrm{st}_n[t]$, $\mathrm{q}_n^\mathrm{tv}[t]$, $\epsilon_n[t]$, $i=1$, and $\nu=0.8$
\While{$\nu<20$}
    \State $\boldsymbol\theta^{i}=\arg\min_{\boldsymbol\theta}\mathrm{E^{apx}_{\nu}}\left({\boldsymbol\theta}\right)$ via gradient-descent search starting from $\boldsymbol\theta^{i-1}$.
    \State ${i:=i+1}$
    \State $\nu:=1.2\nu$
\EndWhile
\end{algorithmic}
\end{algorithm}

\begin{algorithm}[b]
\caption{Gradient-descent algorithm}
\label{alg:gradient}
\begin{algorithmic}[1]
\Require $\mathrm{q^{st}}[t]$, $\mathrm{q^{tv}}[t]$, $\epsilon[t]$, $\nu$, and $j=1$
    \While { $\mathrm{E^{apx}_{\nu}}\left(\boldsymbol\theta^{j-1}\right)-\mathrm{E^{apx}_{\nu}}\left(\boldsymbol\theta^{j}\right)\geq10^{-5}$}
        \State $\Delta{\boldsymbol\theta}:=-\nabla_{\boldsymbol\theta}{\mathrm{E^{apx}_{\nu}}\left({\boldsymbol\theta}^{j}\right)}$
        \While {$\mathrm{E^{apx}_{\nu}}\left({\boldsymbol\theta}^{j}+\omega\Delta{\boldsymbol\theta}\right)>\mathrm{E^{apx}_{\nu}}\left({\boldsymbol\theta}^{j}\right)-0.1\omega||\Delta{\boldsymbol\theta}||_2^2$ or $\rho(\mathbf{f})\geq1$}
        \State $\omega:=0.7\omega$
        \EndWhile
        \State ${\boldsymbol\theta}^{j+1}:={\boldsymbol\theta}^{j}+\omega\Delta{\boldsymbol\theta}$
        \State ${j:=j+1}$
    \EndWhile
\end{algorithmic}
\end{algorithm}

The gradient-descent mechanism in Algorithm~\ref{alg:sysid} is described by Algorithm~\ref{alg:gradient}. The algorithm contains two loops. In the outer loop, $\boldsymbol\theta$ is moved along the direction of negative gradient $-\nabla_{\boldsymbol\theta}\mathrm{E^{apx}_{\nu}}(\boldsymbol\theta)$ with a step-size $\omega$. The loop is terminated when the decrement of the cost function between consecutive loops is less than a small threshold $\delta$. On our database, we found that setting $\delta=10^{-5}$ is sufficient. The inner loop of Algorithm~\ref{alg:gradient} is a standard backtracking line search algorithm (see \cite{Boyd}), which determines an appropriate step-size $\omega$. To calculate the gradient $\nabla_{\boldsymbol\theta}\mathrm{E^{apx}_{\nu}}(\boldsymbol\theta)$, we have
\begin{align}
\label{eq:gradient}
&\nabla_{\boldsymbol\theta}\mathrm{E^{apx}_{\nu}}(\boldsymbol\theta)\\\nonumber
=&\frac{1}{\mathrm{NT}}\sum_{n=1}^{\mathrm N}\sum_{t=1}^{\mathrm T}\left[\left.\diff{\mathrm{U}_{\nu}\left(x,\epsilon_n[t]\right)}{x}\right\vert_{x=\widehat{\mathrm{q}}_n(t,\boldsymbol\theta)-\mathrm{q}^\mathrm{tv}_n[t]}\right]\nabla_{\boldsymbol\theta}\widehat{\mathrm{q}}_n(t,\boldsymbol\theta).
\end{align}
In \eqref{eq:gradient}, $\diff{\mathrm{U}_{\nu}\left(x,\epsilon\right)}{x}$ can be directly derived from \eqref{eq:u_apprx}. The calculation of $\nabla_{\boldsymbol\theta}\widehat{\mathrm{q}}_n(t,\boldsymbol\theta)$ is not straightforward since the dynamic model has a recurrent structure. Specifically, the input-output relationship of the HW model can be written as:
\begin{align}
\label{eq:recurrent_gradient}
\widehat{\mathrm{q}}_n(t,\boldsymbol\theta)=\mathrm{g}\left(\boldsymbol\theta,\big(\widehat{\mathrm{q}}_n\big)_{t-r:t-1},\big(\mathrm{q}^\mathrm{st}_n\big)_{t-r :t}\right),
\end{align}
where the function $\mathrm{g}(\cdot)$ is the combination of \eqref{eq:oe}, \eqref{eq:input}, and \eqref{eq:output}. The model output $\widehat{\mathrm{q}}_n(t,\boldsymbol\theta)$ depends not only on $\boldsymbol\theta$ but also on previous system outputs $\big(\widehat{\mathrm{q}}_n\big)_{t-1:t-r}$, which depend on $\boldsymbol\theta$ as well \footnote{Here $\big(\widehat{\mathrm{q}}_n\big)$ also depends on $\big(\mathrm{q^{st}}\big)$. But in parameter training, $\big(\mathrm{q^{st}}\big)$ is treated as a known constant.}. Denoting by $\theta_{i}$ the ${i^\mathrm{th}}$ component of $\boldsymbol\theta$, differentiating both side of \eqref{eq:recurrent_gradient}, we have
\begin{align}
\label{eq:gradient_rec}
\frac{\partial\widehat{\mathrm{q}}_n(t,\boldsymbol\theta)}{\partial\theta_{i}}=
\frac{\partial\mathrm{g}}{\partial\theta_{i}}+\sum_{d=1}^r\frac{\partial\mathrm{g}}{\partial \widehat{\mathrm{q}}_n(t-{d},\boldsymbol\theta)}\frac{\partial \widehat{\mathrm{q}}_n(t-{d},\boldsymbol\theta)}{\partial\theta_{i}}.
\end{align}
Because of the structure of \eqref{eq:gradient_rec}, computing $\frac{\partial\widehat{\mathrm{q}}_n(t,\boldsymbol\theta)}{\partial\theta_i}$ is equivalent to filtering $\frac{\partial\mathrm{g}}{\partial\theta_{i}}$ by a filter with a transfer function
\begin{align}
\mathrm{H}(z)=\frac{1}{1-\sum_{d=1}^r\frac{\partial\mathrm{g}(\cdot)}{\partial \widehat{\mathrm{q}}_n(t-d,\boldsymbol\theta)}z^{-d}},
\end{align}
If $\boldsymbol\theta$ is not appropriately chosen, the filter $\mathrm{H}(z)$ can be unstable. The computed gradient $\frac{\partial\widehat{\mathrm{q}}_n(t,\boldsymbol\theta)}{\partial\theta_{i}}$ could diverge as $t$ increases. It is proved in Appendix~\ref{app:gradient} that, if the root radius\footnote{The root radius of a polynomial is defined as the maximum radius of its complex roots.} $\rho(\mathbf{f})$ of the polynomial $z^r-\sum_{d=1}^rf_dz^{r-d}$ is less than 1, the filter $\mathrm{H}(z)$ is stable. Therefore, in the line search step Algorithm~\ref{alg:gradient}, the step-size $\omega$ is always chosen to be small enough such that the condition $\rho(\mathbf{f})<1$ is satisfied (see line 3-5 in Algorithm \ref{alg:gradient}). For further details about the calculation of $\frac{\partial\widehat{\mathrm{q}}_n(t,\boldsymbol\theta)}{\partial\theta_{i}}$, see Appendix~\ref{app:gradient}.

\subsection{Model Order Selection}
\label{sec:model_order}
Using Algorithm~\ref{alg:sysid}, the optimal parameter $\boldsymbol\theta$ for a given model order $r$ can be determined. This section discusses how to select the model order. First, a possible range of model orders is estimated by inspecting the correlation between the input and output of the HW model, i.e., $\left(\mathrm{q^{st}}\right)_{1:\mathrm{T}}$ and $\left(\mathrm{q^{tv}}\right)_{1:\mathrm{T}}$. Then, the model order is determined in the estimated range using the principle of Minimum Description Length.

\begin{figure}[!t]
\centering
\subfigure[Lipschitz quotient]{
\includegraphics[width=0.9\columnwidth]{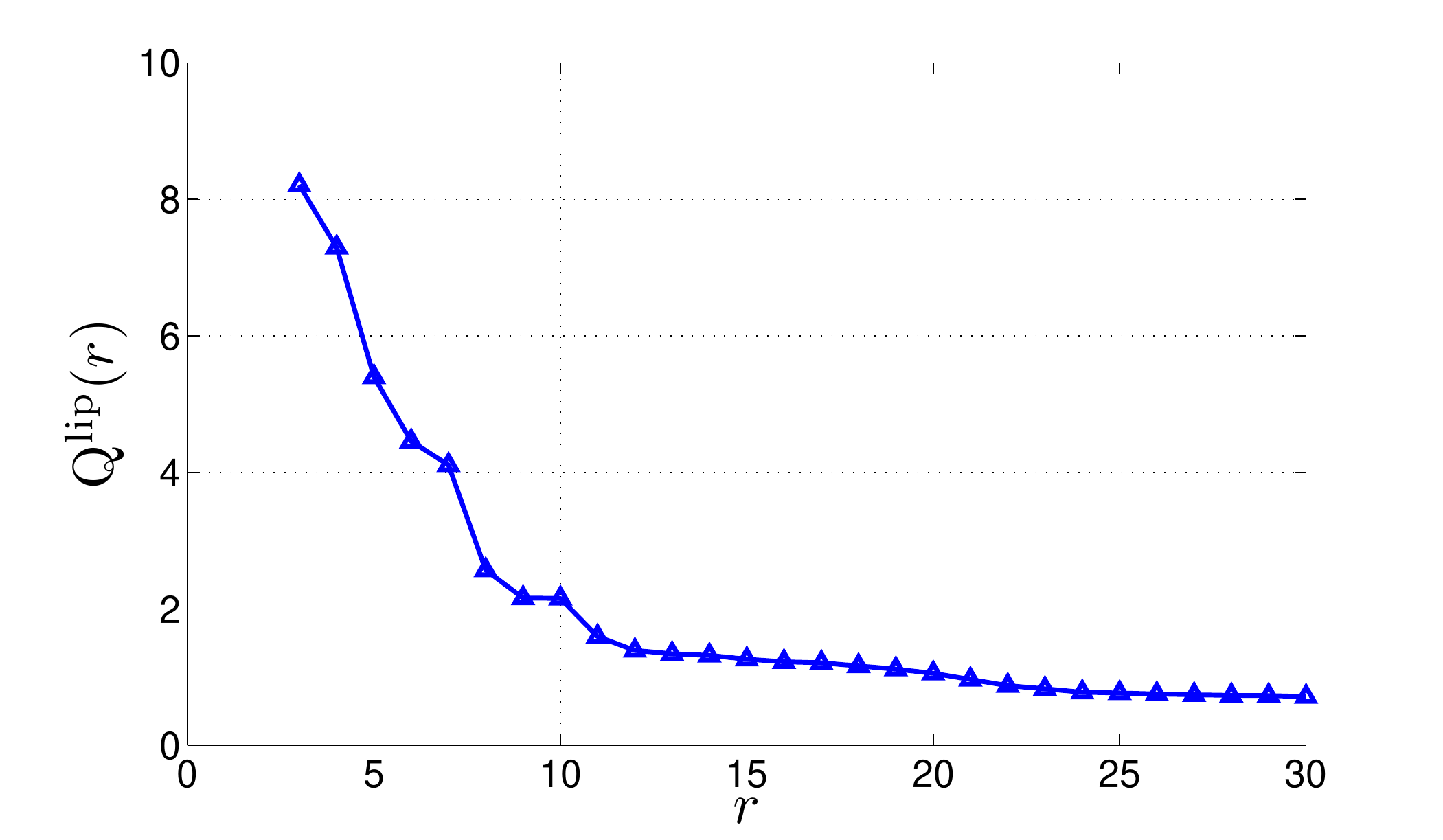}
\label{fig:lip}
}
\subfigure[Description length]{
\includegraphics[width=0.9\columnwidth]{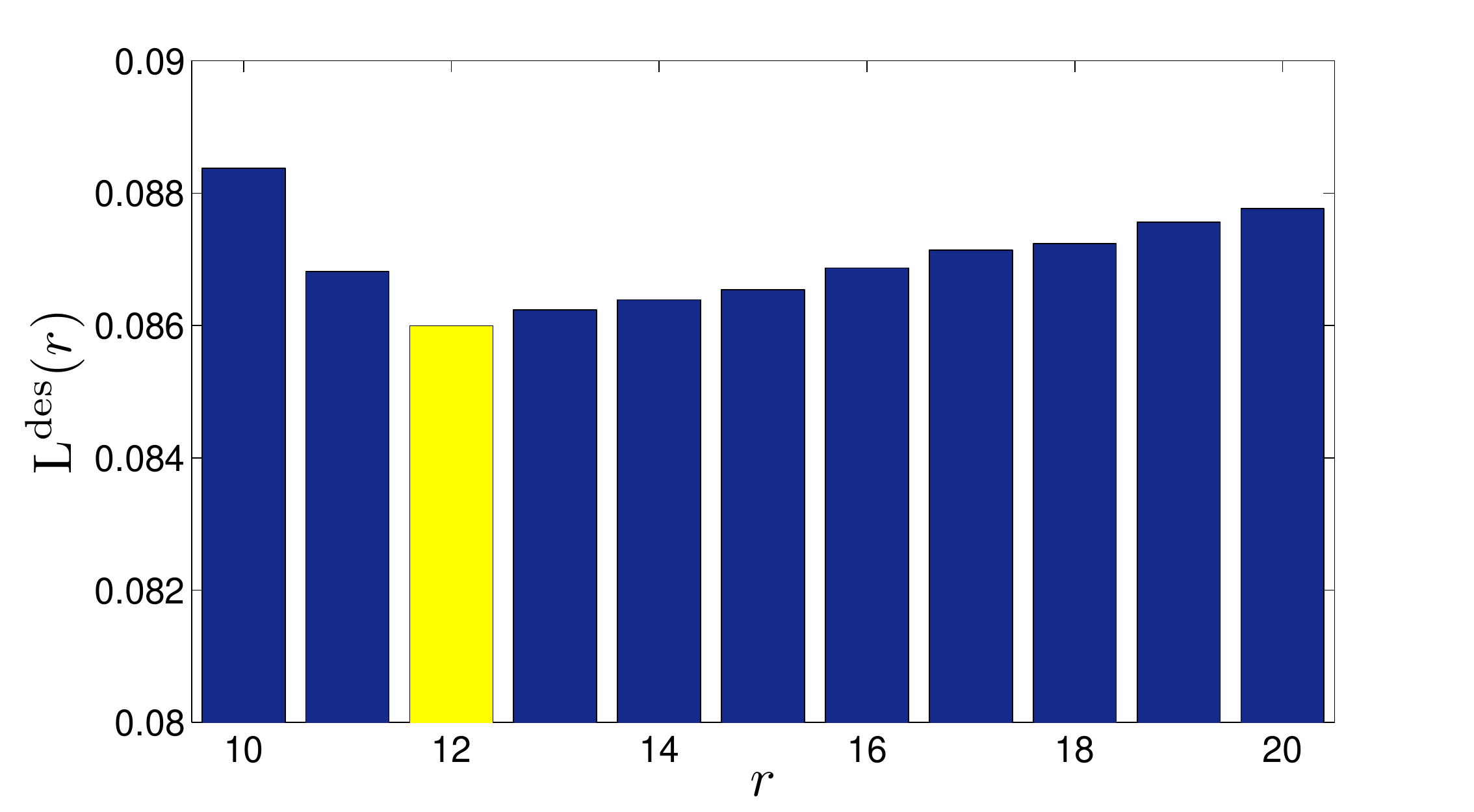}
\label{fig:des}
}
\caption[Optional caption for list of figures]{
Model order selection via (a) Lipschitz quotient and (b) Description length.
}
\label{fig:order}
\end{figure}

The TVSQ at any time depends on the previous viewing experience. In the proposed TVSQ model \eqref{eq:recurrent_gradient}, $\boldsymbol\phi_{r}[t]=\left(\left(\mathrm{q^{st}}\right)^\mathsf{T}_{t-r:t},\left(\mathrm{q^{tv}}\right)^\mathsf{T}_{t-r:t-1}\right)^\mathsf{T}$ has been employed as the model input to capture the previous viewing experience. Thus, identifying the model order $r$ is essentially estimating how much previous viewing experience is relevant to the current TVSQ. In \cite{lipschitz_quotient}, the Lipschitz quotient was proposed to quantify the relevance of $\boldsymbol\phi_{r}$ by
\begin{align}
\mathrm{Q}^\mathrm{lip}(r)=\max_{1\leq t_1<t_2\leq \mathrm{T}}\left(\frac{\left|\mathrm{q^{tv}}[t_1]-\mathrm{q^{tv}}[t_2]\right|}{||\boldsymbol\phi_{r}[t_1]-\boldsymbol\phi_{r}[t_2]||_2}\right).
\end{align} A large $\mathrm{Q}^\mathrm{lip}(r)$ implies that a small change in $\boldsymbol\phi_{r}$ could cause a significant change in $\mathrm{q^{tv}}$ and thus $\boldsymbol\phi_{r}$ is relevant to TVSQ. Conversely, if $\mathrm{Q}^\mathrm{lip}(r)$ is small, the model order $r$ may be larger than necessary. Using $\mathrm{Q}^\mathrm{lip}(r)$, the necessary model order can be estimated. In Fig.~\ref{fig:lip}, the Lipschitz quotients for different values of $r$ are plotted. It can be seen that, as the model order increases, the corresponding Lipschitz quotient decreases significantly when $r$ is less than 10. This means the viewing experience over the previous 10 seconds is closely related to the TVSQ. Therefore, the model order $r$ should be at least 10.

According to the parameterizations of the HW model in \eqref{eq:oe}, \eqref{eq:input}, and \eqref{eq:output}, models of lower order are special cases of the model of higher order. Therefore, in principle, the higher the order, the better performance can be achieve by the model. A large model order, however, may result in over-fitting the model to the training dataset. To select an appropriate order for the HW model, we employed the Minimum Description Length (MDL) criterion, which is widely used in the realm of system identification \cite{MDL}\cite{Sysid}. The description length of an $r$-order model is defined in \cite{Sysid} as
\begin{align}
\label{eq:mdl}
\mathrm{L}^\mathrm{des}(r)=\mathrm{E}(\boldsymbol\theta^*_r)\left(1+(2r+1)\frac{\log(N(\mathrm {T}-r))}{N(\mathrm{T}-r)}\right),
\end{align}
where $\boldsymbol\theta^*_r$ is the model parameter of the $r$-order model determined through Algorithm~\ref{alg:sysid}. The first multiplicative term in \eqref{eq:mdl}, which is defined in \eqref{eq:or} as the outage rate, represents the ability of a model to describe the data. The second multiplicative term increases with the number of parameters $(2r+1)$ and decreases with the size of training set $N(T-r)$. Thus, this term roughly indicate the whether the training set is sufficiently large for training a $r$-order model. The definition of \eqref{eq:mdl} balances the accuracy and the complexity of the model. In Fig.~\ref{fig:des}, the description lengths of the proposed models under different model orders are plotted. It is seen that the minimum description length is achieved at $r=12$. Therefore, $r=12$ were selected.

\section{Model Evaluation and Analysis}
\label{sec:model_analysis}
In this section, the efficacy of the proposed HW model is studied first. Then, four important properties of the proposed model were studied. They are the impact of the initial state, the stability for online TVSQ prediction, the input and output nonlinearities, and the impulse response of the IIR filter.
\begin{figure*}
\centering
\subfigure[Video \#1]{
\includegraphics[width=0.31\columnwidth]{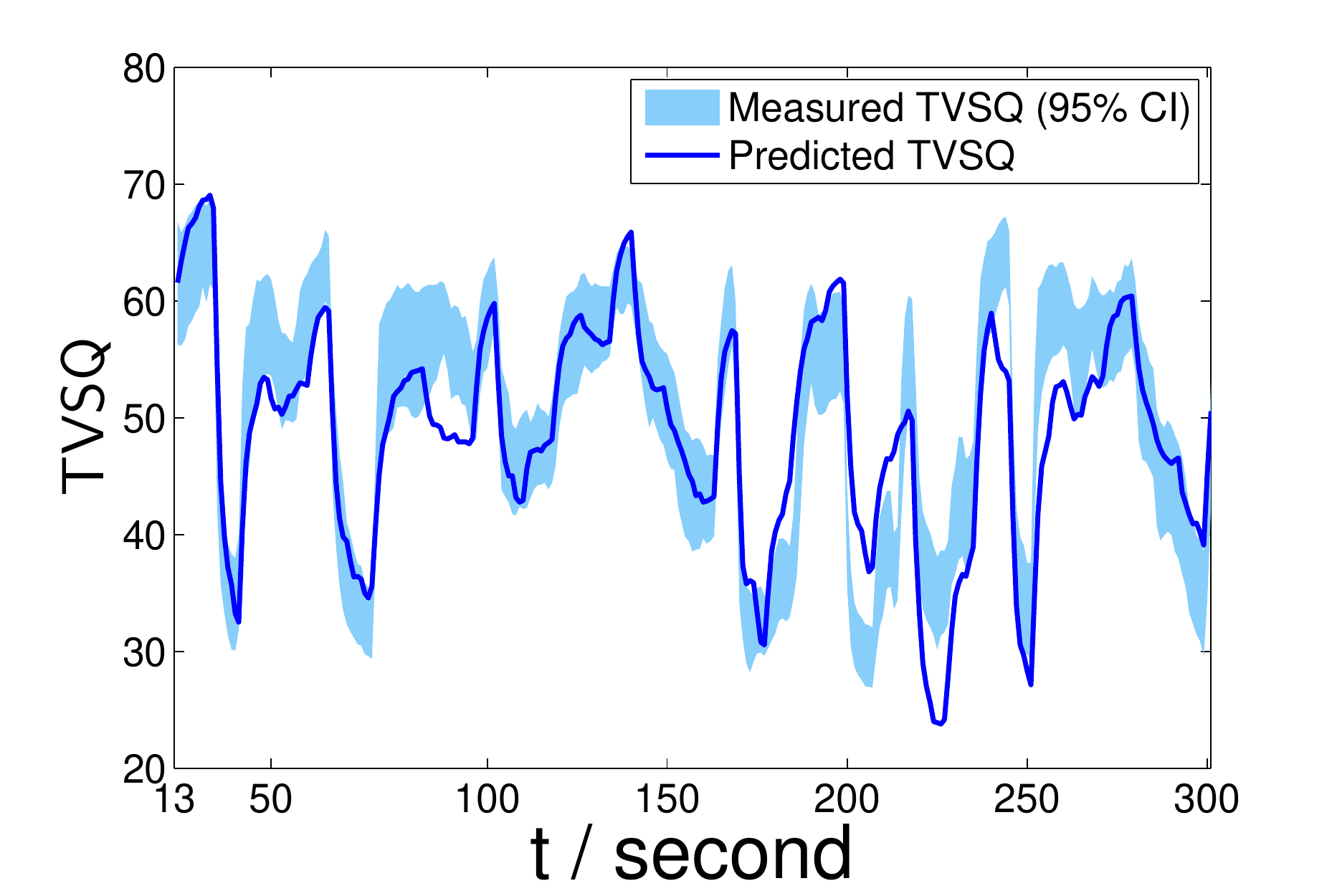}
\label{fig:seq1}
}
\subfigure[Video \#2]{
\includegraphics[width=0.31\columnwidth]{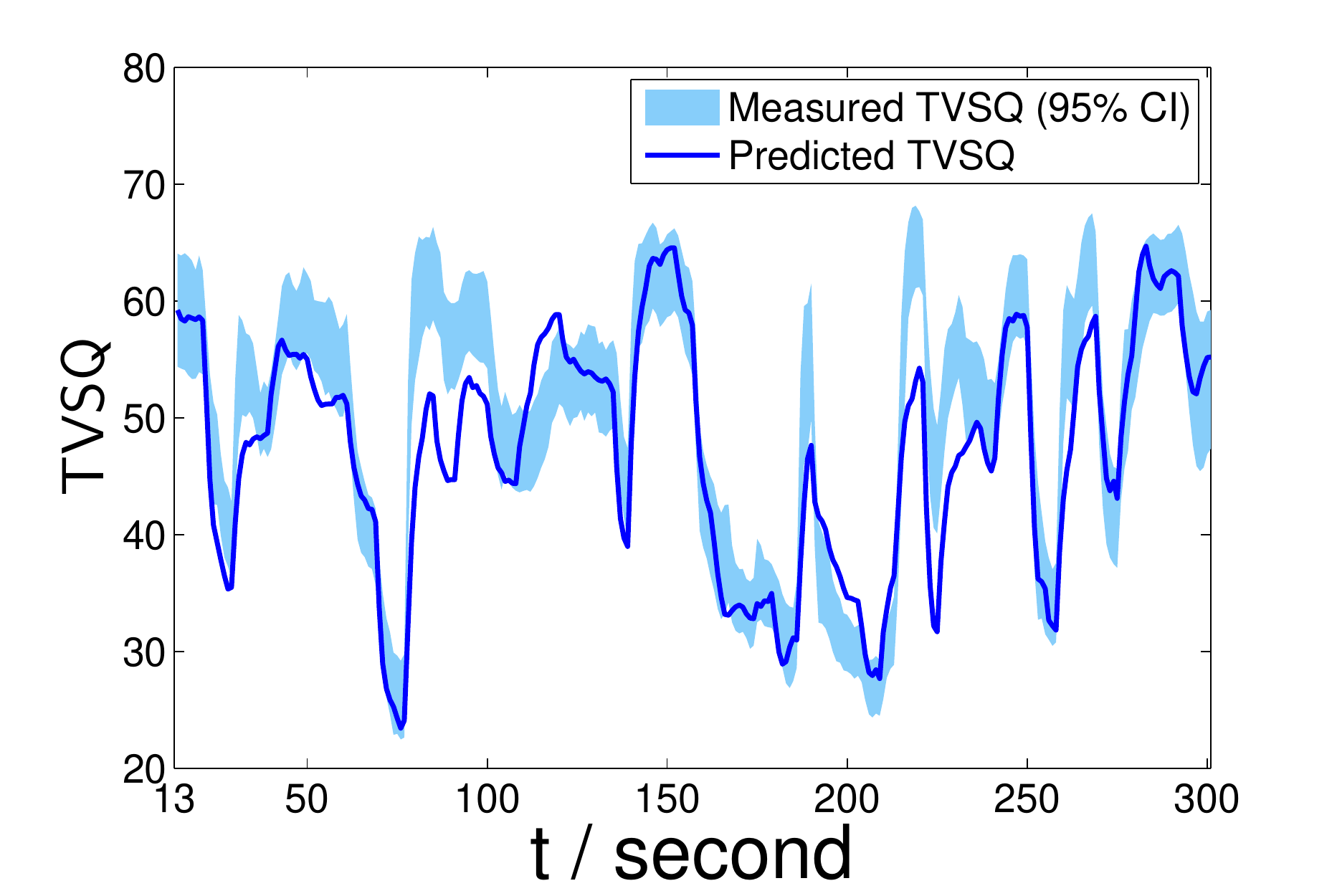}
\label{fig:seq2}
}
\subfigure[Video \#3]{
\includegraphics[width=0.31\columnwidth]{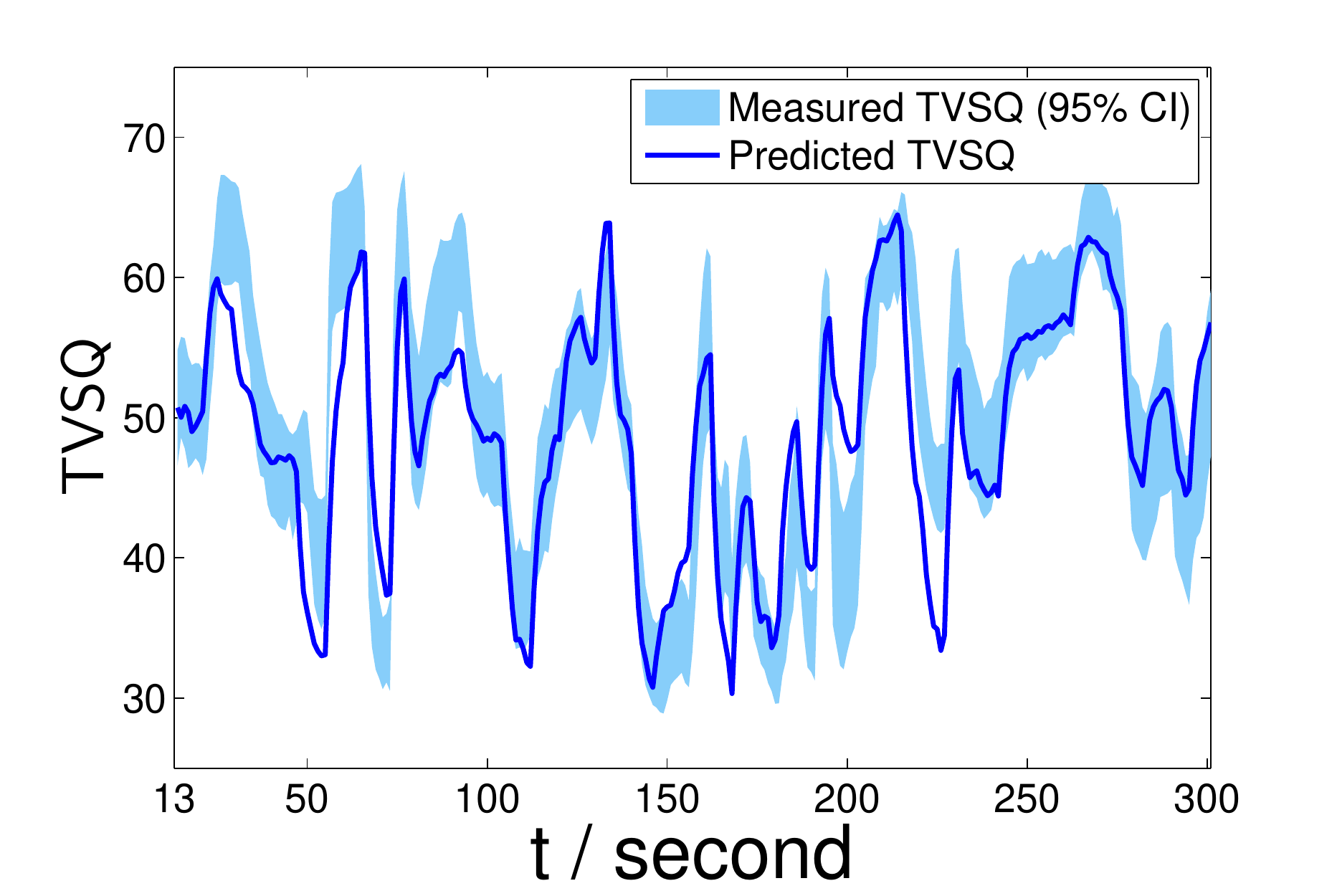}
\label{fig:seq3}
}
\subfigure[Video \#4]{
\includegraphics[width=0.31\columnwidth]{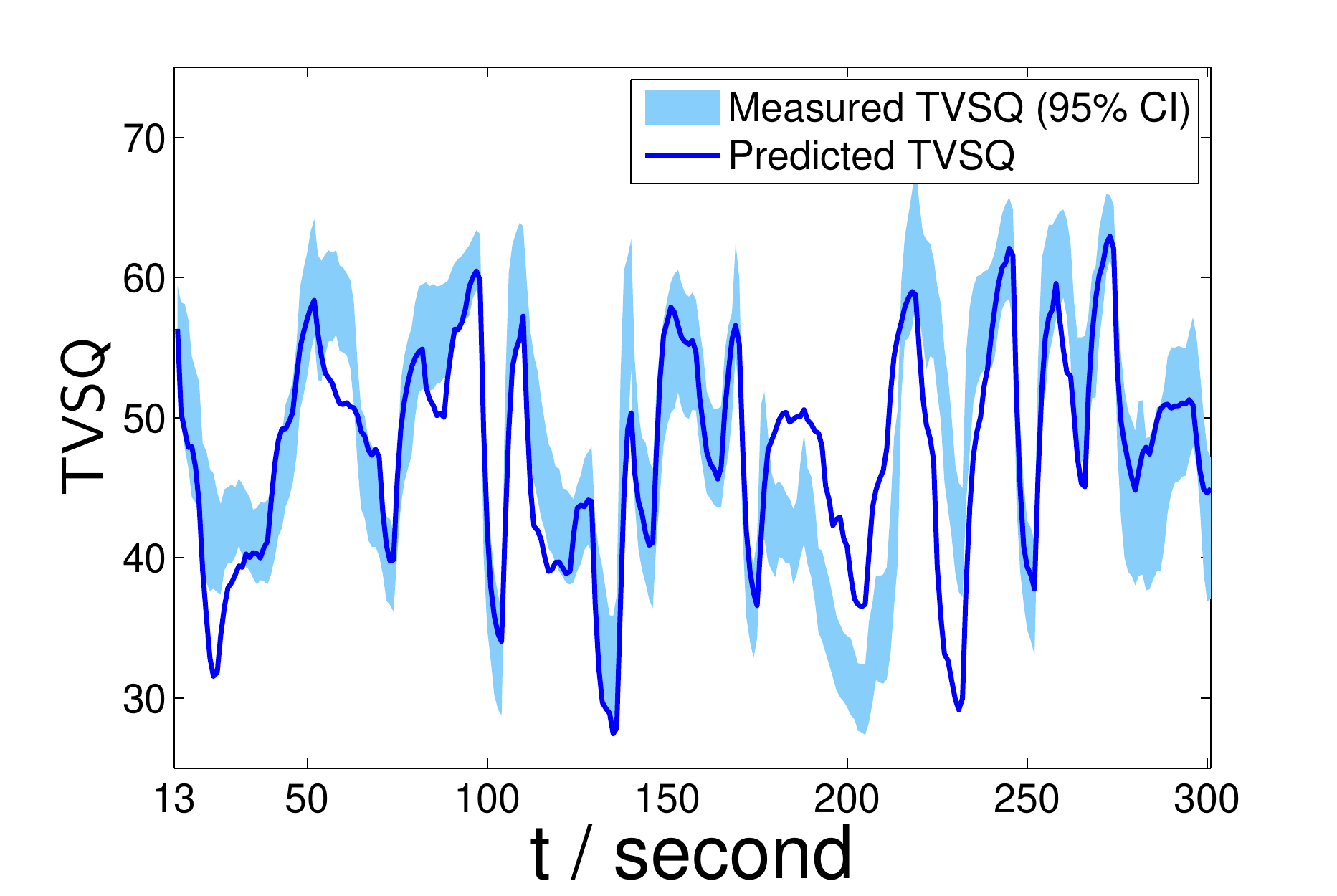}
\label{fig:seq4}
}
\subfigure[Video \#5]{
\includegraphics[width=0.31\columnwidth]{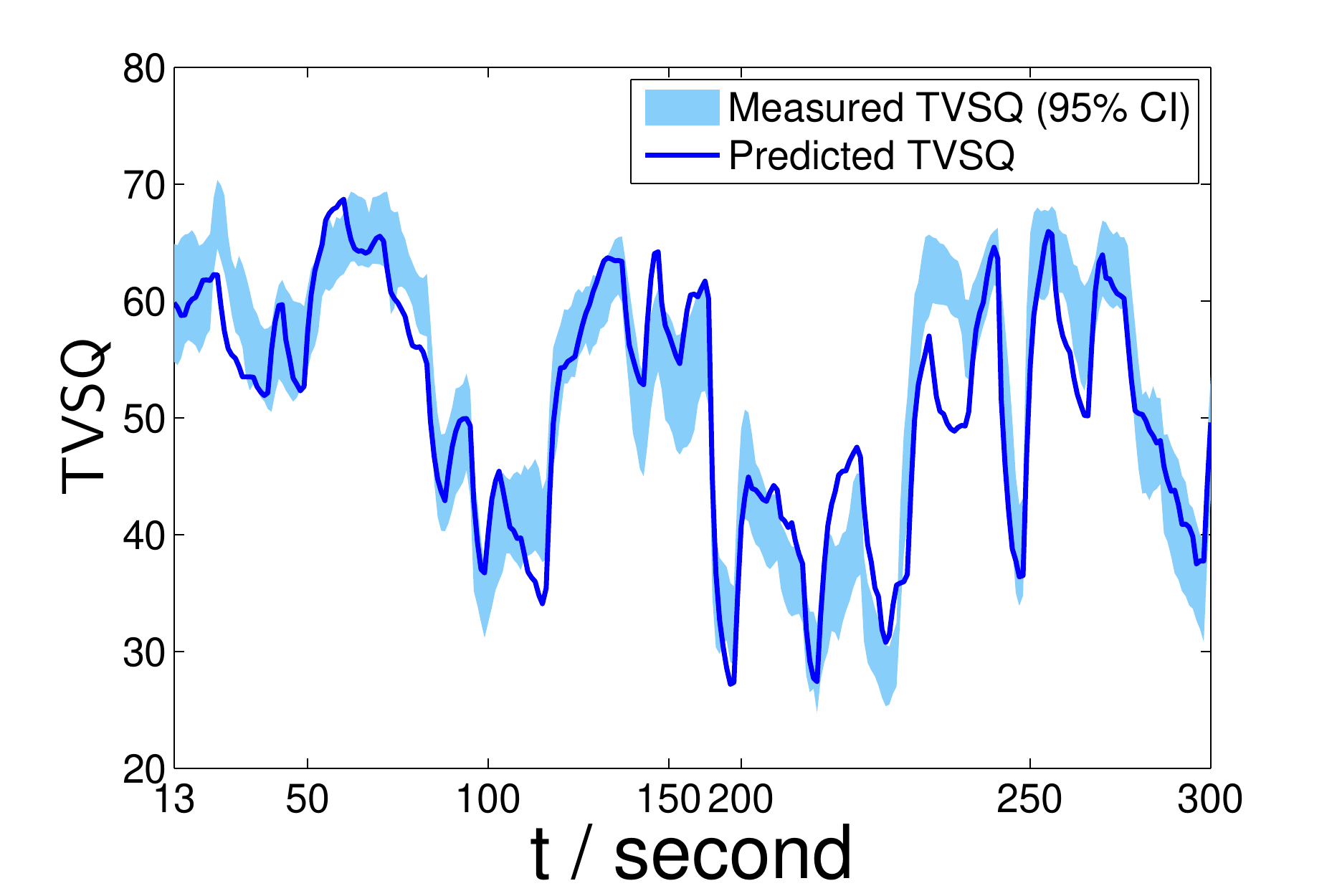}
\label{fig:seq5}
}
\subfigure[Video \#6]{
\includegraphics[width=0.31\columnwidth]{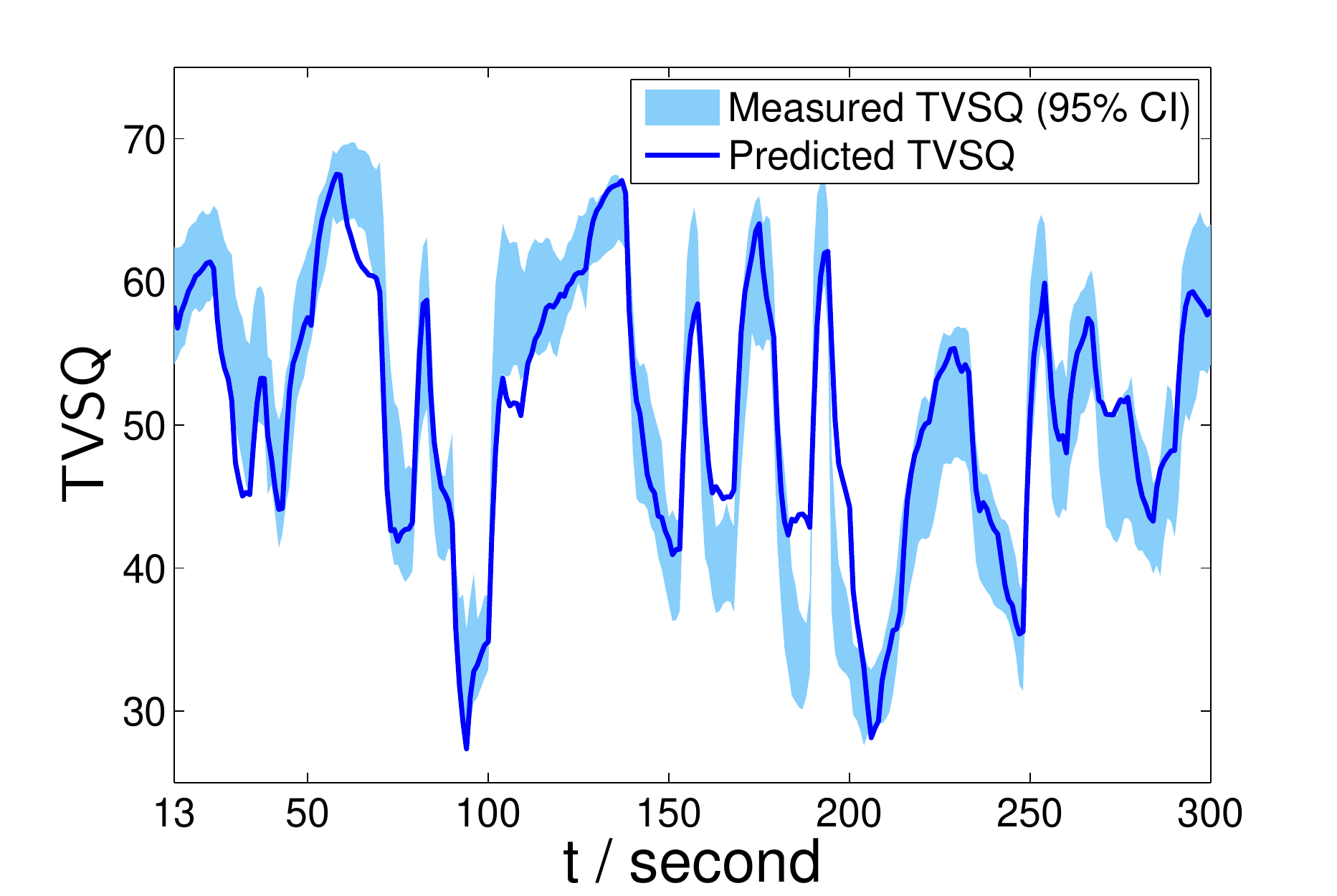}
\label{fig:seq6}
}
\subfigure[Video \#7]{
\includegraphics[width=0.31\columnwidth]{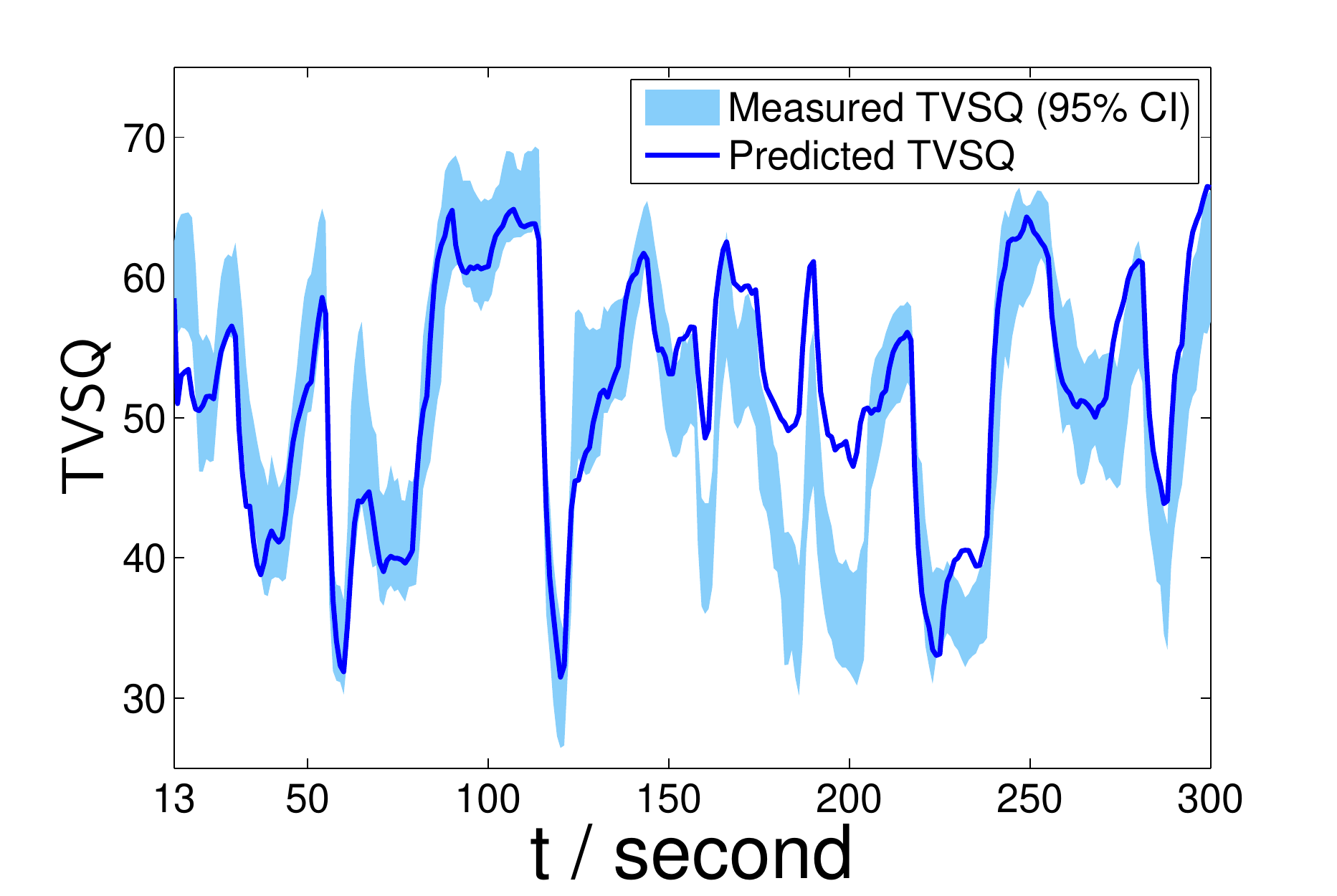}
\label{fig:seq7}
}
\subfigure[Video \#8]{
\includegraphics[width=0.31\columnwidth]{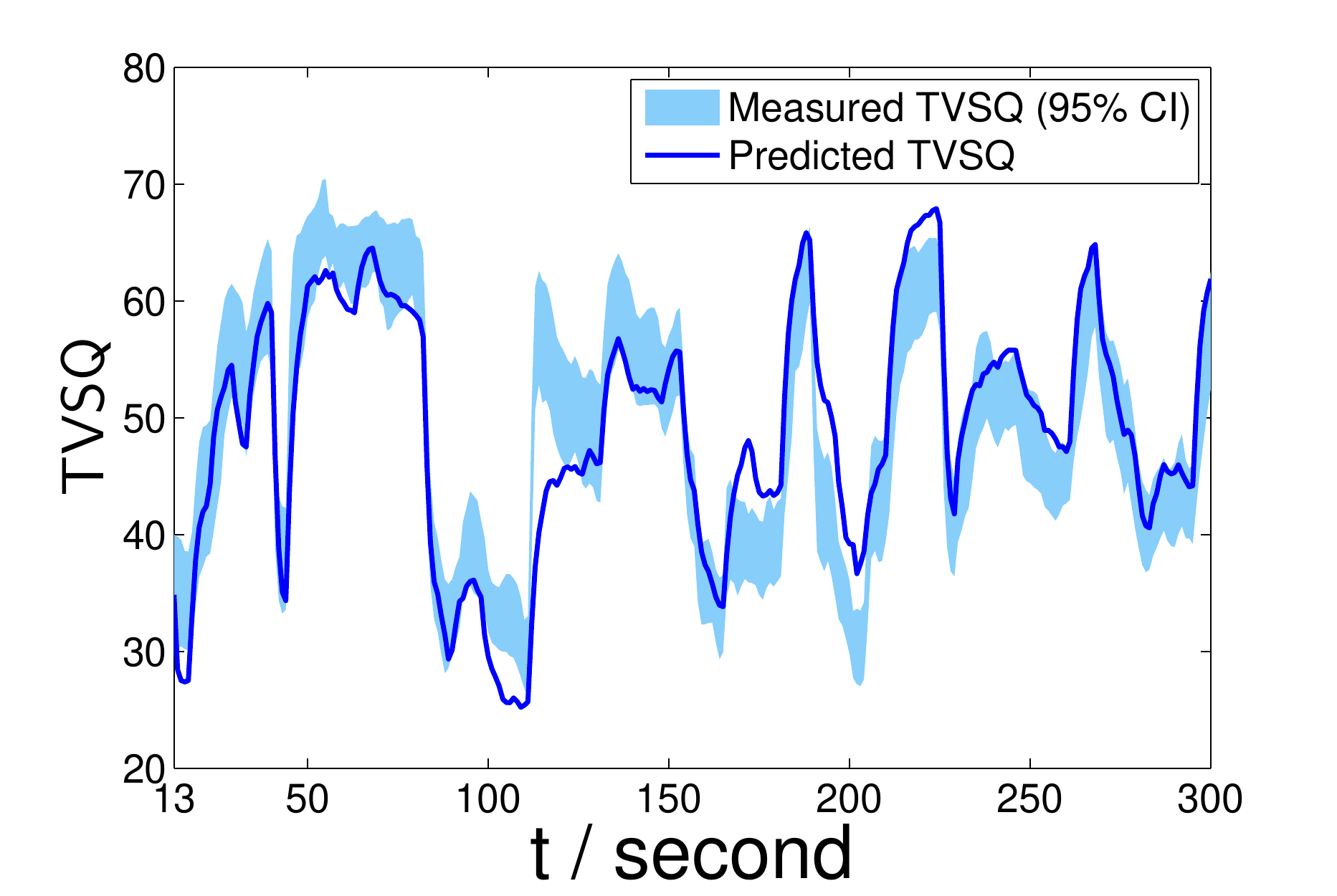}
\label{fig:seq8}
}
\subfigure[Video \#9]{
\includegraphics[width=0.31\columnwidth]{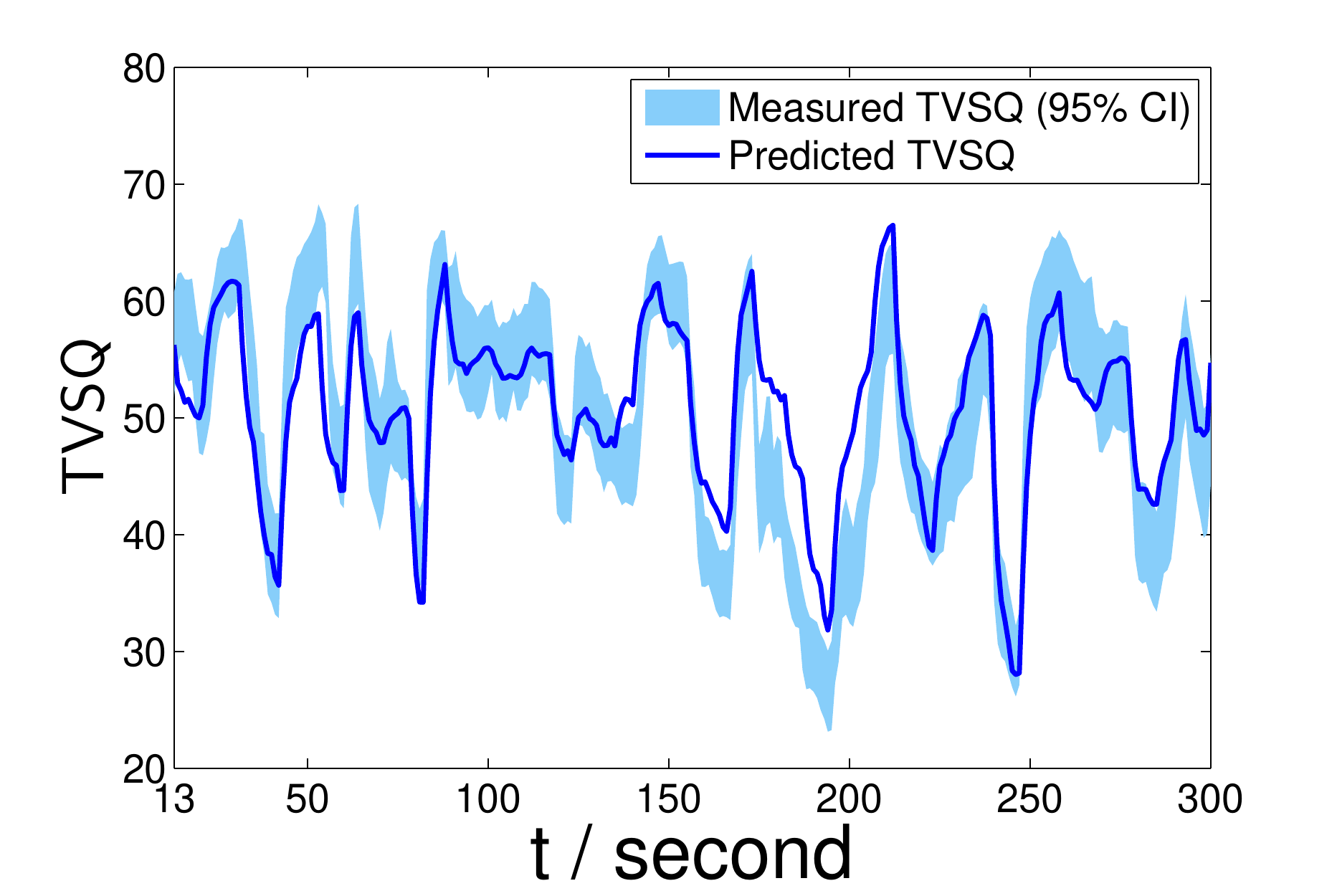}
\label{fig:seq9}
}
\subfigure[Video \#10]{
\includegraphics[width=0.31\columnwidth]{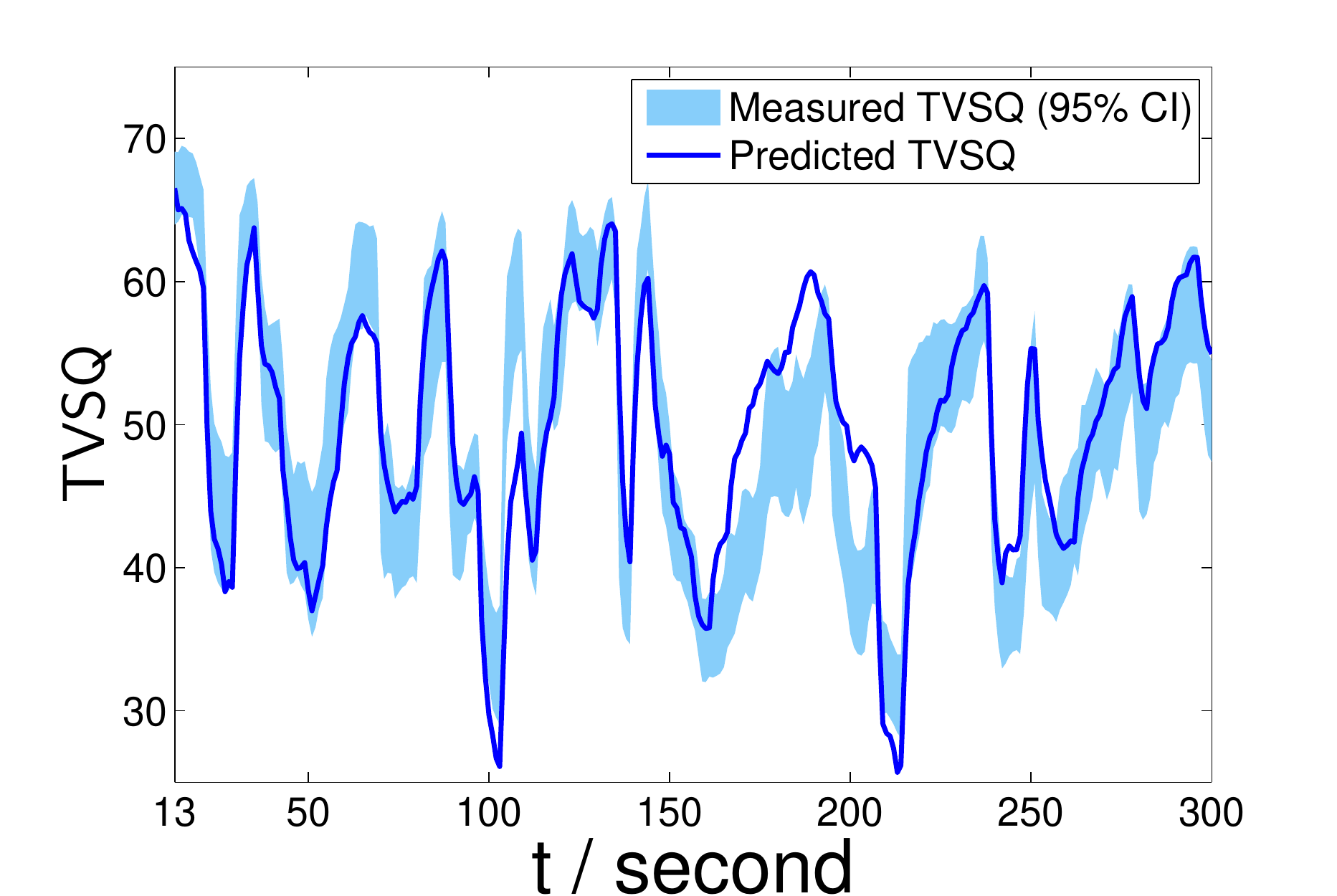}
\label{fig:seq10}
}
\subfigure[Video \#11]{
\includegraphics[width=0.31\columnwidth]{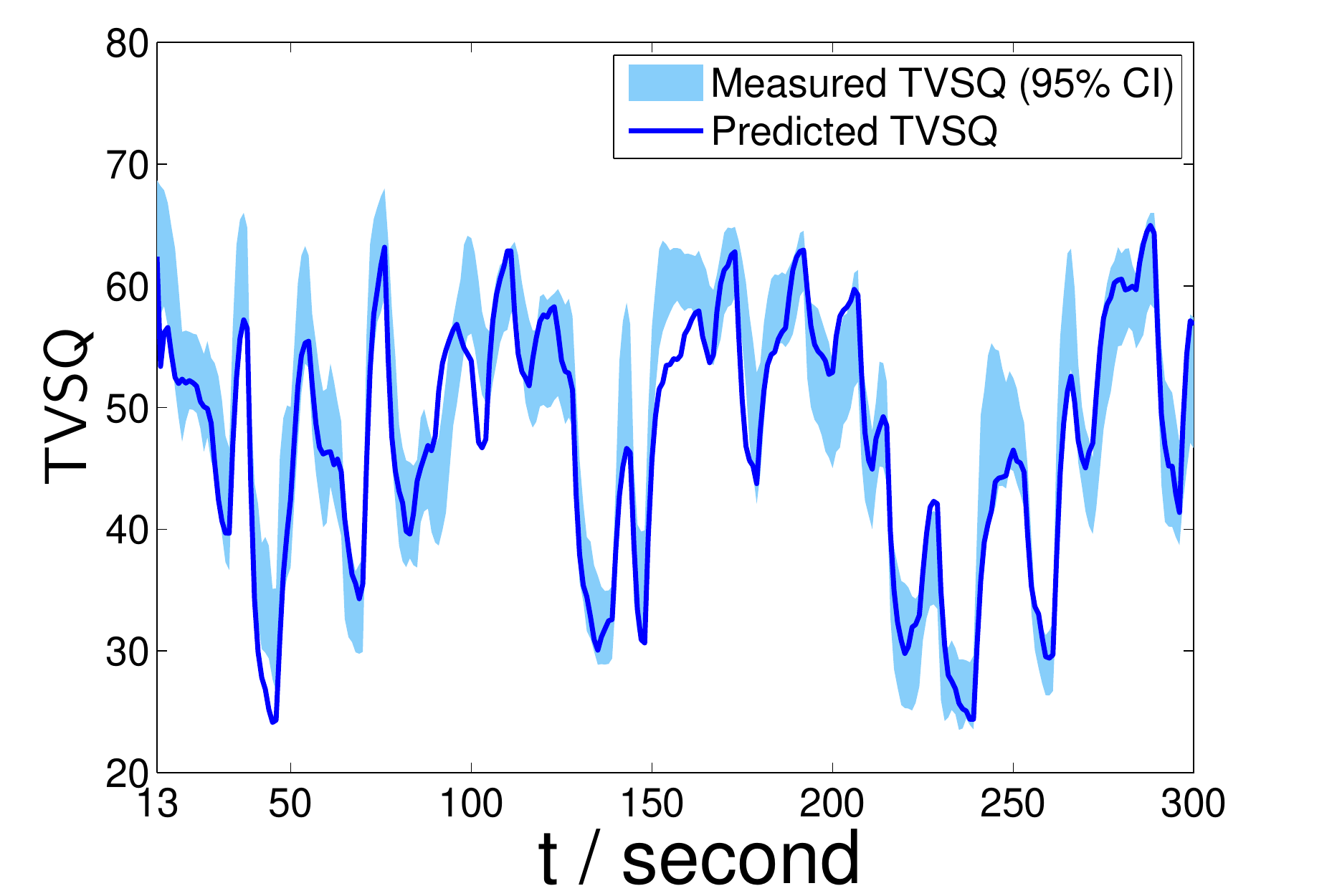}
\label{fig:seq11}
}
\subfigure[Video \#12]{
\includegraphics[width=0.31\columnwidth]{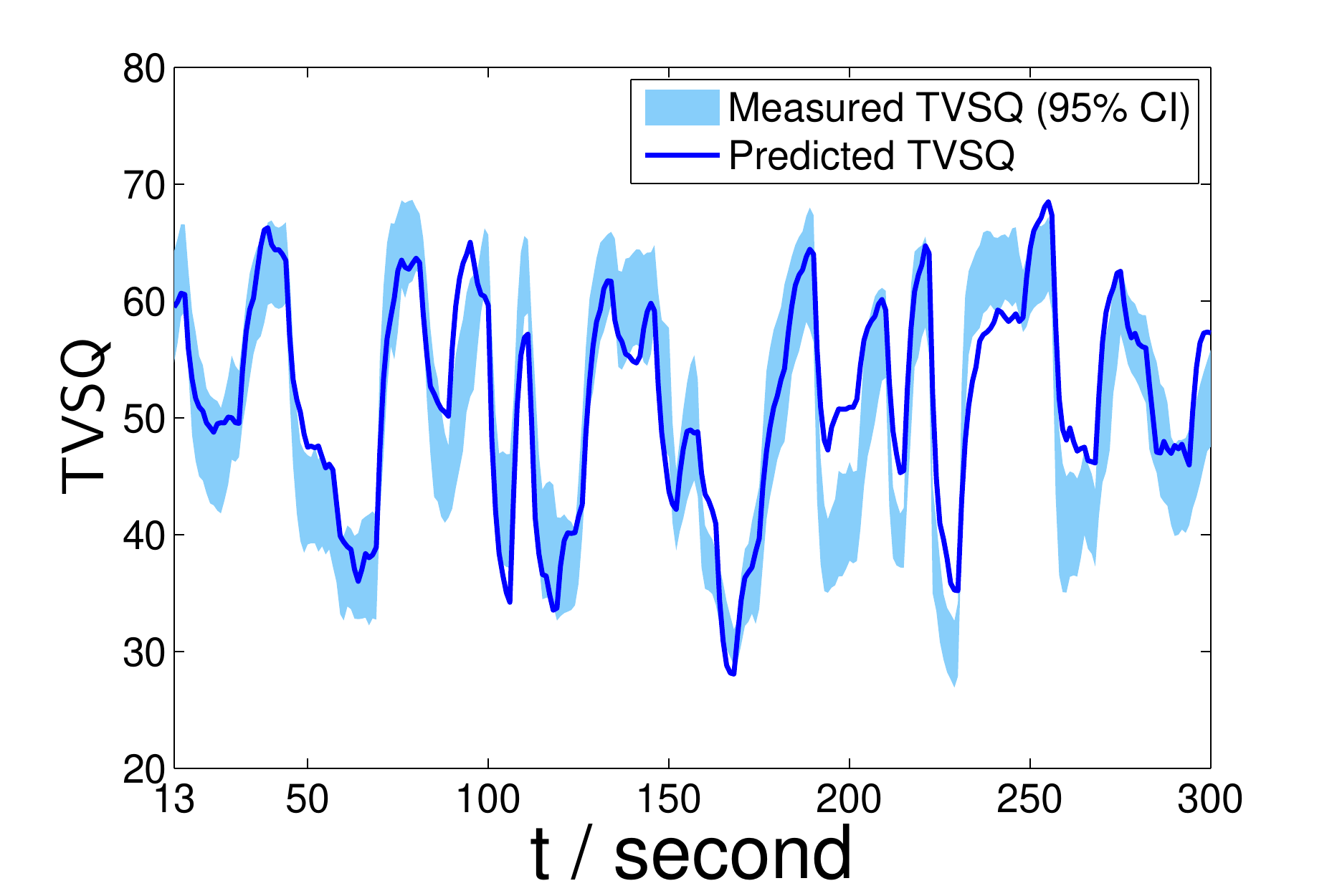}
\label{fig:seq12}
}
\subfigure[Video \#13]{
\includegraphics[width=0.31\columnwidth]{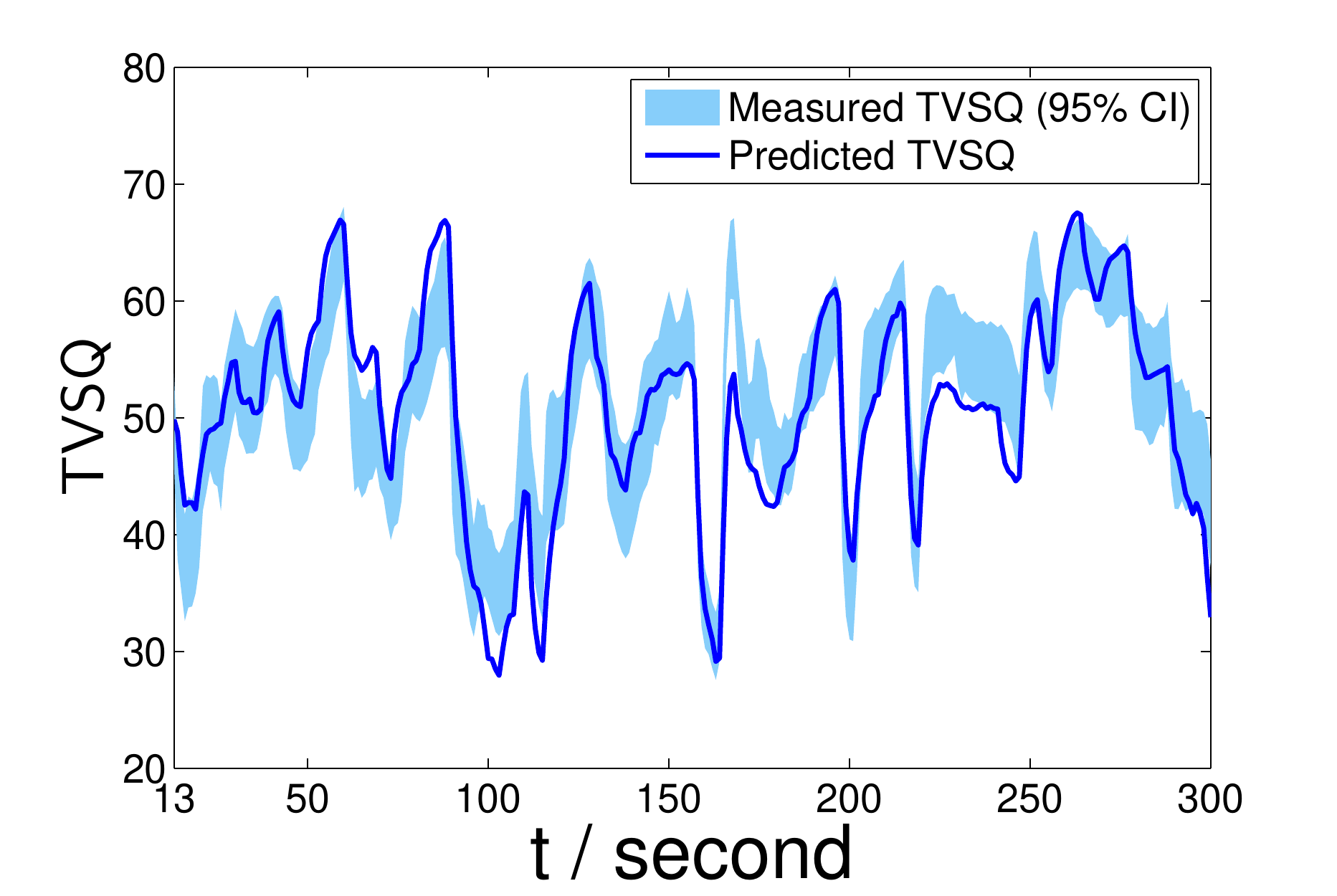}
\label{fig:seq13}
}
\subfigure[Video \#14]{
\includegraphics[width=0.31\columnwidth]{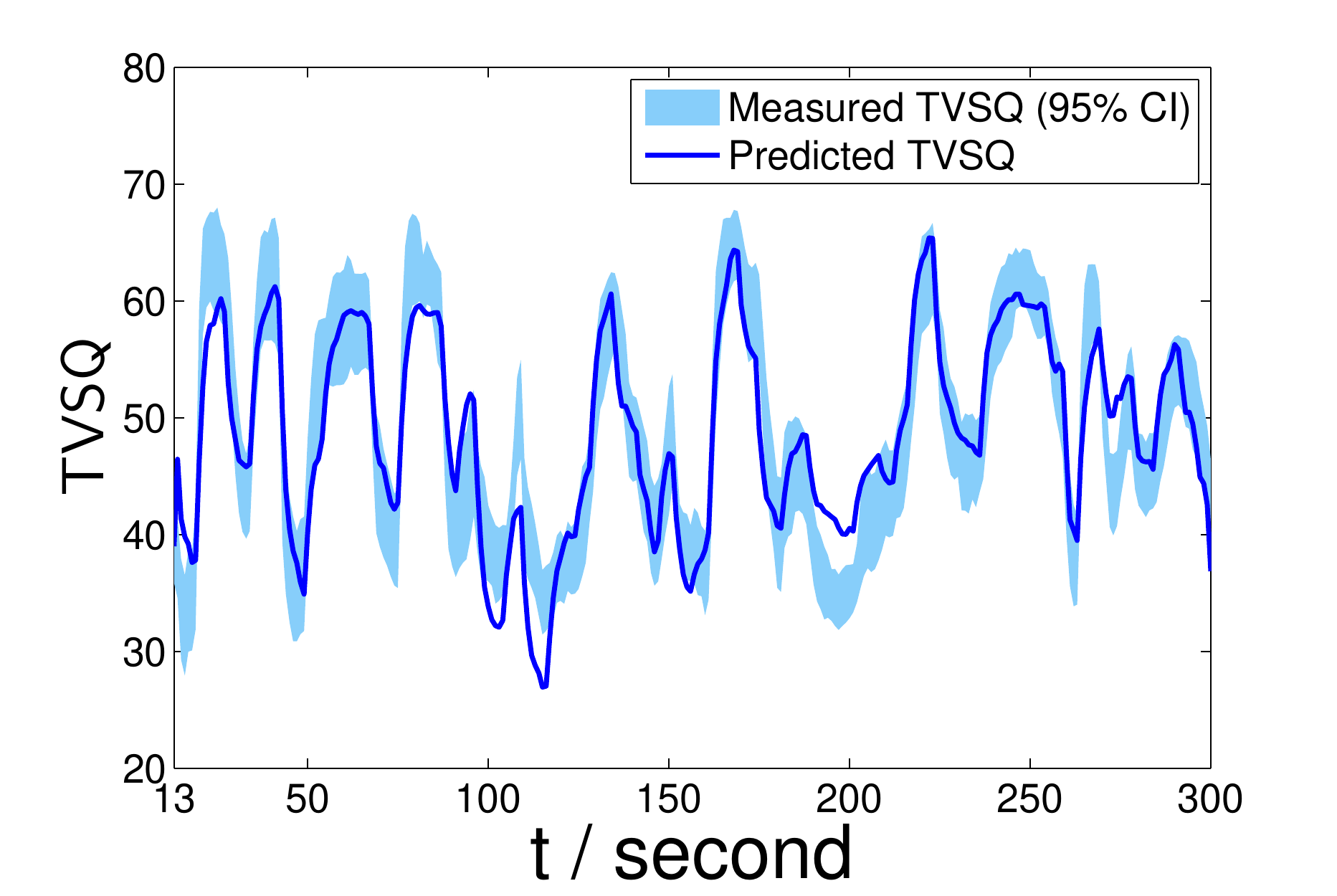}
\label{fig:seq14}
}
\subfigure[Video \#15]{
\includegraphics[width=0.31\columnwidth]{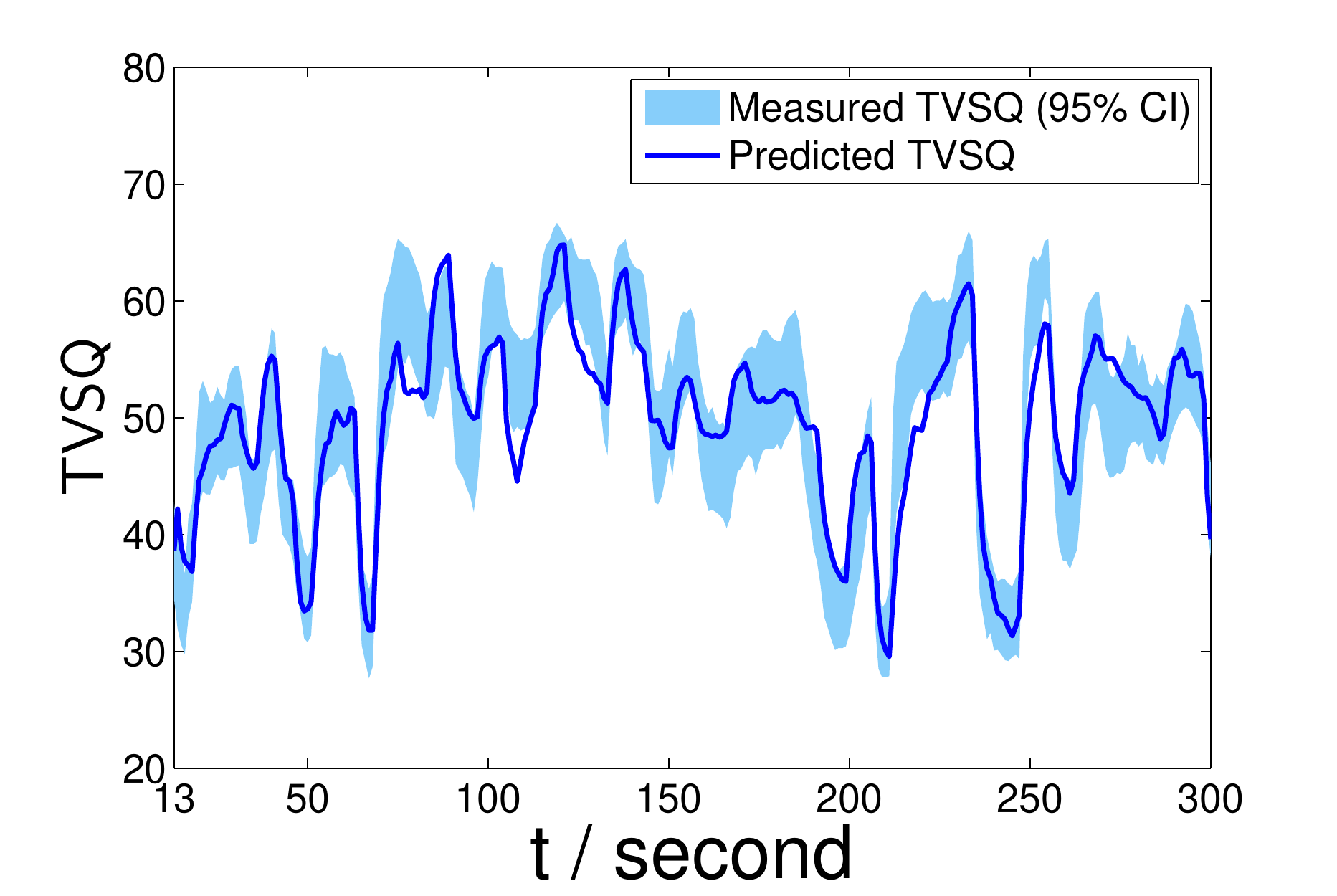}
\label{fig:seq15}
}
\caption[Optional caption for list of figures]{
The predicted TVSQ and the 95\% confidence interval (CI) of the TVSQs measured in the subjective study. Since the Hammerstein-Wiener model predicts TVSQs using the STSQs of previous 12 seconds, the plots start from $t=13$.
}
\label{fig:performance}
\end{figure*}
\subsection{Model Evaluation and Validation}
\label{sec:validation}
The model parameters were trained using our database via Algorithm \ref{alg:sysid}.
Table~\ref{tab:val} list the outage rate of the trained model on all of the 15 test videos. The average outage rate is 8.06\%. This means that the model can accurately predict 91.94\% of the TVSQs in the database. Furthermore, Table~\ref{tab:val} also list the linear correlation coefficient and the Spearman's rank correlation coefficient between the predicted TVSQ and the measured TVSQ values. The average linear correlation and rank correlation achieved by our model is 0.885 and 0.880, respectively.
In Fig.~\ref{fig:performance}, the predicted TVSQs and the 95\% confidence interval of the measured TVSQs are plotted. The proposed model effectively tracked the measured TVSQs of the 15 quality-varying videos.

\begin{center}
\begin{table*}[ht]
\caption{Performance of the proposed model on the database}
\label{tab:val}
\centering
\resizebox{\columnwidth}{!}{%
\begin{tabular}{c||l|l|l|l|l|l|l|l|l|l|l|l|l|l|l|l}
    \hline
    &\#1&\#2&\#3&\#4&\#5&\#6&\#7&\#8&\#9&\#10&\#11&\#12&\#13&\#14&\#15&mean\\
    \hline
    outage rate(\%)&12.15 &11.46 &9.38 &18.06 &9.72 &4.17 & 10.76& 8.33&8.33 &8.33 & 3.82&7.64 & 1.74&6.25 &0.69 &{\bf 8.06}\\
    \hline
    linear correlation&0.868 & 0.897& 0.862& 0.785& 0.919& 0.936& 0.859& 0.896& 0.845& 0.863& 0.938& 0.898& 0.892& 0.916& 0.906&{\bf 0.885}\\
    \hline
    rank correlation&0.881&0.857 &0.875 & 0.814&0.897 & 0.943& 0.872& 0.901& 0.833& 0.859& 0.911& 0.899& 0.870& 0.927& 0.866&{\bf 0.880}\\
    \hline
\end{tabular}%
}
\end{table*}
\end{center}

In the proposed method, the TVSQ is estimated by the Hammerstein-Wiener model using the RRED-predicted STSQs of the previous twelve seconds. In Table \ref{tab:pool}, the proposed method is compared with several basic pooling methods, i.e., the maximum, the minimum, the median, and the mean of the RRED-predicted STSQs in the previous twelve seconds. It is seen that the proposed method achieves a significantly lower outage rate and a much stronger correlation with the measured TVSQs.
\begin{center}
\begin{table*}[ht]
\caption{Performance comparison with different TVSQ pooling methods}
\centering
\begin{tabular}{c||l|l|l|l|l}
    \hline
    &max&min&median&mean&proposed\\
    \hline
    outage rate(\%)&34.26&32.06&26.76&22.22&{\bf 8.06}\\
    \hline
    linear correlation&0.497&0.541&0.589&0.702&{\bf 0.885}\\
    \hline
    rank correlation&0.475&0.515&0.611&0.693&{\bf 0.880}\\
    \hline
\end{tabular}%
\label{tab:pool}
\end{table*}
\end{center}

Table~\ref{tab:stsq_predictor} shows the performance of the proposed TVSQ prediction method when the STSQ predictor is PSNR, MS-SSIM, and RRED. It may be seen that the RRED-based model outperforms both the MS-SSIM-based model and the PSNR-based model. This can be attributed to the high accuracy of RRED in STSQ prediction. It can also be observed that the performance of the MS-SSIM-based model is close to that of RRED. Since MS-SSIM has lower complexity, it may be attractive as a low-complexity alternative to RRED in the TVSQ prediction model if slightly lower prediction accuracy is acceptable.

\begin{center}
\begin{table*}[ht]
\caption{Performance of the TVSQ prediction model with different STSQ predictors}
\centering
\begin{tabular}{c||l|l|l}
    \hline
    STSQ predictors &PSNR&MS-SSIM&RRED\\
    \hline
    outage rate(\%)&21.8&11.5&{\bf 8.06}\\
    \hline
    linear correlation&0.754&0.855&{\bf 0.885}\\
    \hline
    rank correlation&0.744&0.862&{\bf 0.880}\\
    \hline
\end{tabular}%
\label{tab:stsq_predictor}
\end{table*}
\end{center}

To rule out the risk of over-fitting the model to the TVSQ database, a leave-one-out cross-validation protocol were employed to check whether the model trained on our database is robust. Each time, the 5 videos corresponding to the same reference video were selected as the validation set and trained the model parameters on the other 10 videos. This procedure was repeated such that all the videos are included once in the validation set. The results are summarized in Table~\ref{tab:cross_val}. Comparing with the models trained on the whole database, the performance of the models in the cross-validation is only slightly degraded. Therefore, the model obtained from our database appears to be robust.


\begin{center}
\begin{table*}[ht]
\caption{Results of leave-one-out cross-validations. Here \{$n_1$,...,$n_2$\} denotes the set of video sequences with sequence numbers from $n_1$ to $n_2$. The performance of the models obtained in cross-validation is shown in boldface. The performance of the model that is trained on the whole database is also listed for comparison.}
\centering
\resizebox{\columnwidth}{!}{%
\begin{tabularx}{1.2\textwidth}{c||Y|Y||Y|Y||Y|Y}
\hline
validation set &\multicolumn{2}{ c||}{\{1,...,5\}}&\multicolumn{2}{ c||}{\{6,...,10\}}&\multicolumn{2}{ c}{\{11,...,15\}}\\
\hline
training set&\{1,...,15\}&\{6,...,15\}&\{1,...,15\}&\{1,...,5,11,...,15\}&\{1,...,15\}&\{1,...,11\}\\
\hline
outage rate (\%)&12.154&{\bf 13.75}&7.98&{\bf10.00}&4.03&{\bf5.00}\\
\hline
linear correlation&0.866&{\bf0.860}&0.881&{\bf0.875}&0.910&{\bf0.903}\\
\hline
rank correlation&0.864&{\bf0.862}&0.882&{\bf0.879}&0.895&{\bf0.889}\\
\hline
\end{tabularx}%
}
\label{tab:cross_val}
\end{table*}
\end{center}

\subsection{Impact of Initial State}
\label{sec:ini_state}
\begin{figure}[!b]
\centering
\includegraphics[width=0.9\columnwidth]{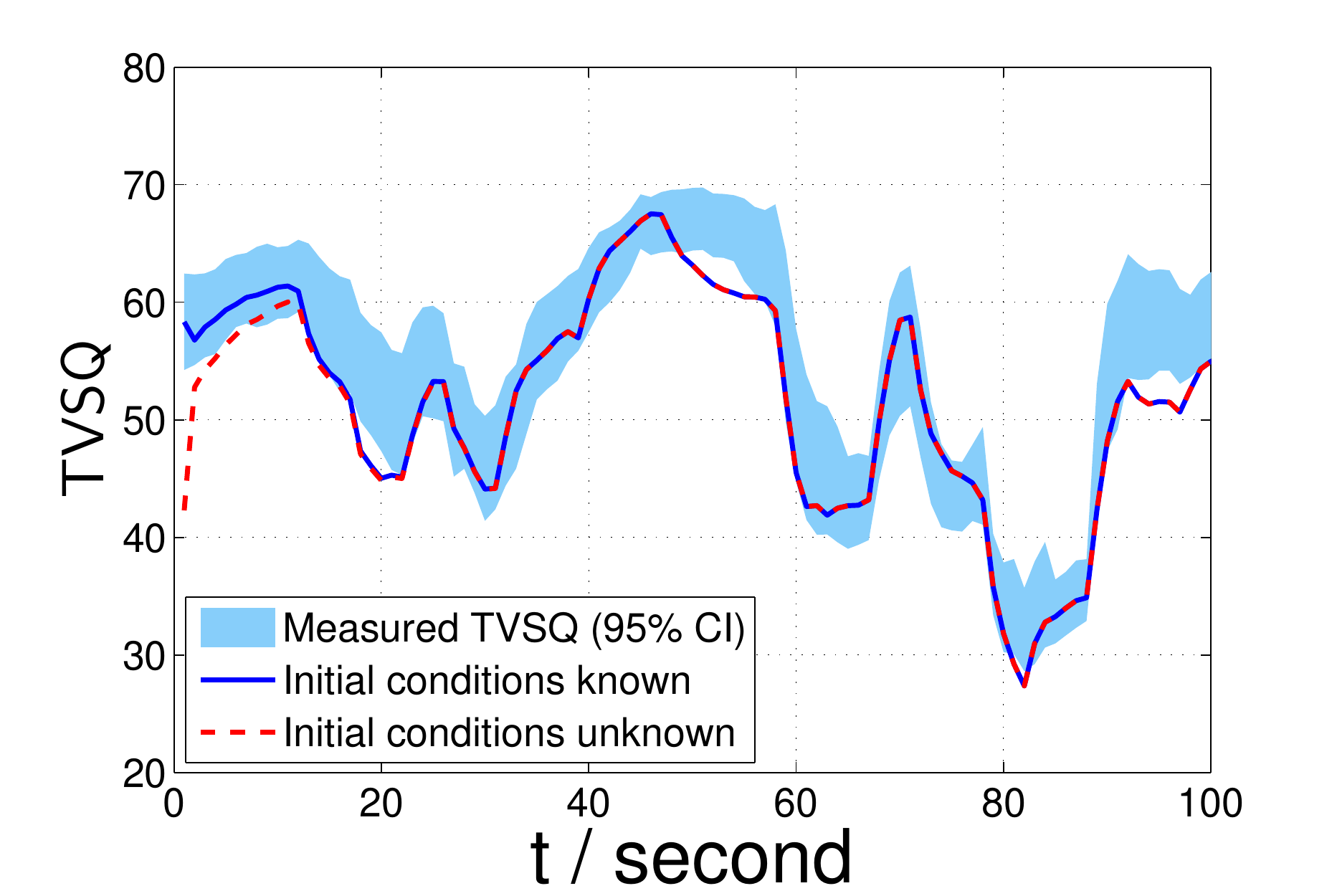}
\caption[Optional caption for list of figures]{
An illustration of the impact of initial state on predicted TVSQ. Dashed Line: Initial condition $\left(\mathrm{v}\right)_\mathrm{1:r}$ is set to be zero. Solid Line: Initial condition is assumed to be known, i.e., $\left(\mathrm{v}\right)_{1:r}=\left(\mathrm{k^{-1}_{\boldsymbol\gamma}}(\mathrm{q^{tv}})\right)_{1:r}$, where $\left(\mathrm{q^{tv}})\right)_{1:r}$ is the measured TVSQ in the subjective study.
}
\label{fig:ini}
\end{figure}
As indicated in Section~\ref{sec:Modelintro}, the initial conditions $\big(\mathrm{v}\big)_{1:r}$ are required to estimate TVSQ. For online video streaming applications, however, $\big(\mathrm{v}\big)_{1:r}$ is unavailable because $\big(\mathrm{v}\big)_{1:r}$ is given by $\big(\mathrm{q^{tv}}\big)_{1:r}$ and the latter is the TVSQ of the first $r$ seconds of the video. This section studies the impact of the unavailability of the initial conditions.

The transfer function of the linear filter is
\begin{equation}
\mathrm{H}(z)=\frac{\sum_{d=0}^rb_{d}z^{-d}}{1-\sum_{d=1}^rf_{d}z^{-d}}.
\end{equation}
According to classical results from system theory, if the root radius of the denominator polynomial $z^r-\sum_\mathrm{d=1}^rf_{d}z^{r-d}$, is less than 1, the impact of the initial condition fades to 0 as $t\rightarrow\infty$ exponentially fast. Denoting by $\rho(\mathbf{f})$ the root radius of $z^r-\sum_{d=1}^rf_{d}z^{r-d}$, the fading speed is $\rho(\mathbf{f})^t$. Here, we define the quantity $\tau(\mathbf{f})=-3/\ln\rho(\mathbf{f})$. Over every $\tau(\mathbf{f})$ seconds, the impact of the initial state fades to $e^{-3}\approx5\%$ of its original level. Therefore, $\tau(\mathbf{f})$ indicates the delay before our TVSQ model starts to track TVSQ. For the model trained on the TVSQ database, $\tau(\mathbf{f})= 15.1895$ seconds. This means that our model cannot accurately predict the TVSQs of the first 15.1895 seconds of the video. For quality monitoring of long videos, this delay is tolerable. In Fig.~\ref{fig:ini}, the impact of the initial state on one of the quality-varying videos is illustrated. The figure shows the predicted TVSQ when the initial state $\big(\mathrm{v}\big)_{1:r}$ is simply set to zero. For comparison, the predicted TVSQs when the initial state is assumed to be perfectly known is also shown in the figure. It can be seen that the predicted TVSQs in both cases coincide with each other when $t>15$ seconds. It also justifies that, for long videos, the impact of the initial condition diminishes over time.

\subsection{Stability for Online TVSQ Prediction}
\label{sec:BIBO}
The goal of our TVSQ model is for the online TVSQ prediction. Different from our video database, where each video is 5 minutes long, the videos streamed over HTTP can be much longer. Therefore, it is necessary to check the long-term stability of the proposed model. Specifically, for any quality-varying video, the estimated TVSQ should be bounded within the RDMOS scale of $[0,100]$. Since the filter is a linear system, we have $\big(\mathrm{v}\big)_\mathrm{1:T}=\big(\mathrm{h}\big)_\mathrm{1:T}*\big(\mathrm{u}\big)_\mathrm{1:T}$, where $\mathrm{h}[t]$ is the impulse response of the linear filter. It is well-known that $||\mathbf{v}||_\infty=||\mathbf{h}||_1||\mathbf{u}||_\infty$, i.e., that the dynamic range of $\mathrm{v}[t]$ is a dilation of the dynamic range of $\mathrm{v}[t]$. For our TVSQ model, we found that $||\mathbf{h}||_1\approx0.3853$. Given that the dynamic range of $\mathrm{q^{st}}[t]$ is $[0,100]$, then the dynamic range of $\widehat{\mathrm{q}}[t]$ is found to be $[10.2661, 78.9525]$. Therefore, the proposed model has bounded output in RDMOS scale.

\subsection{The Input and Output Nonlinearities}
\begin{figure}
\centering
\subfigure[Input nonlinearity]{
\includegraphics[width=0.9\columnwidth]{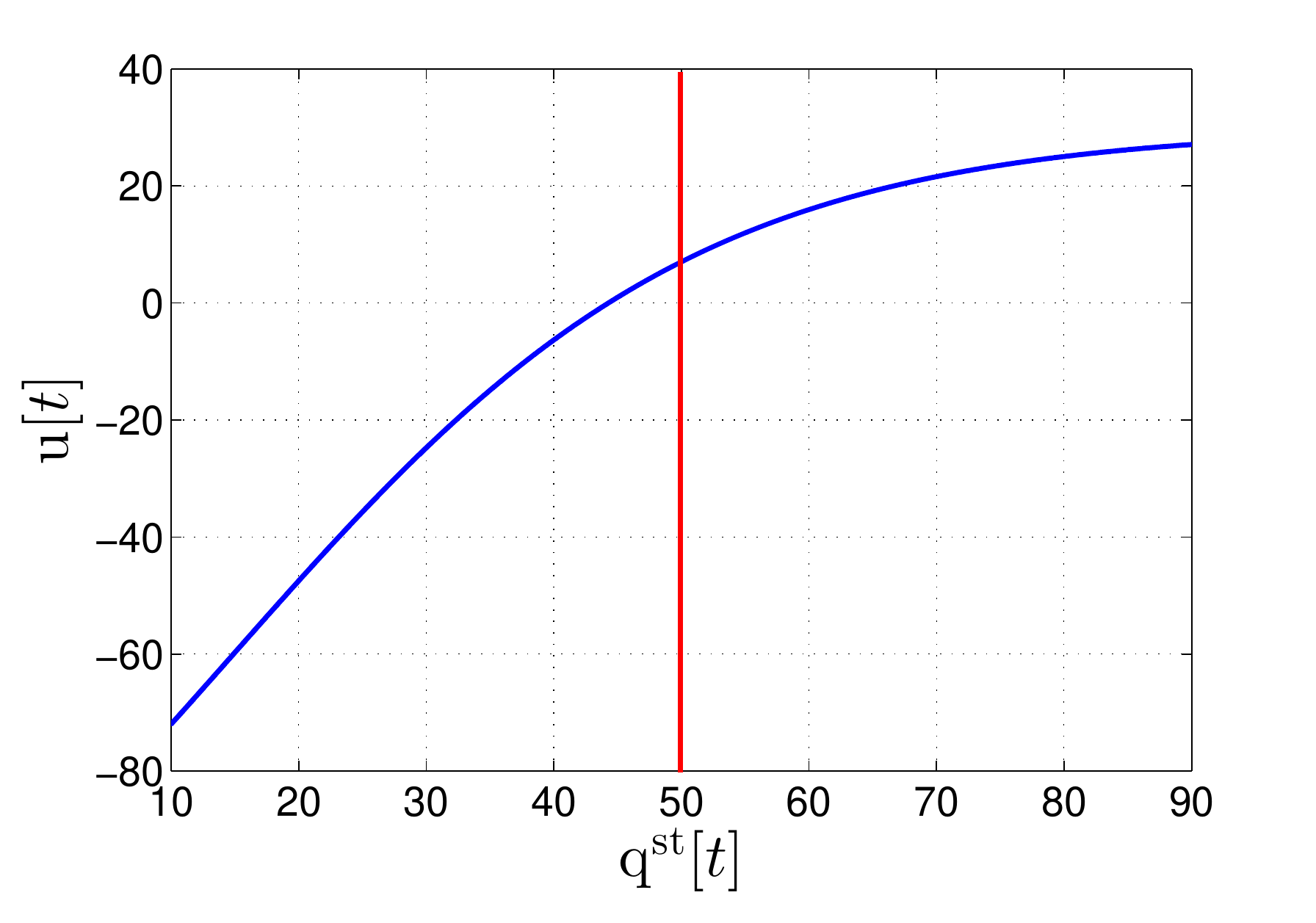}
\label{fig:input}
}
\hfil
\subfigure[Output nonlinearity]{
\includegraphics[width=0.9\columnwidth]{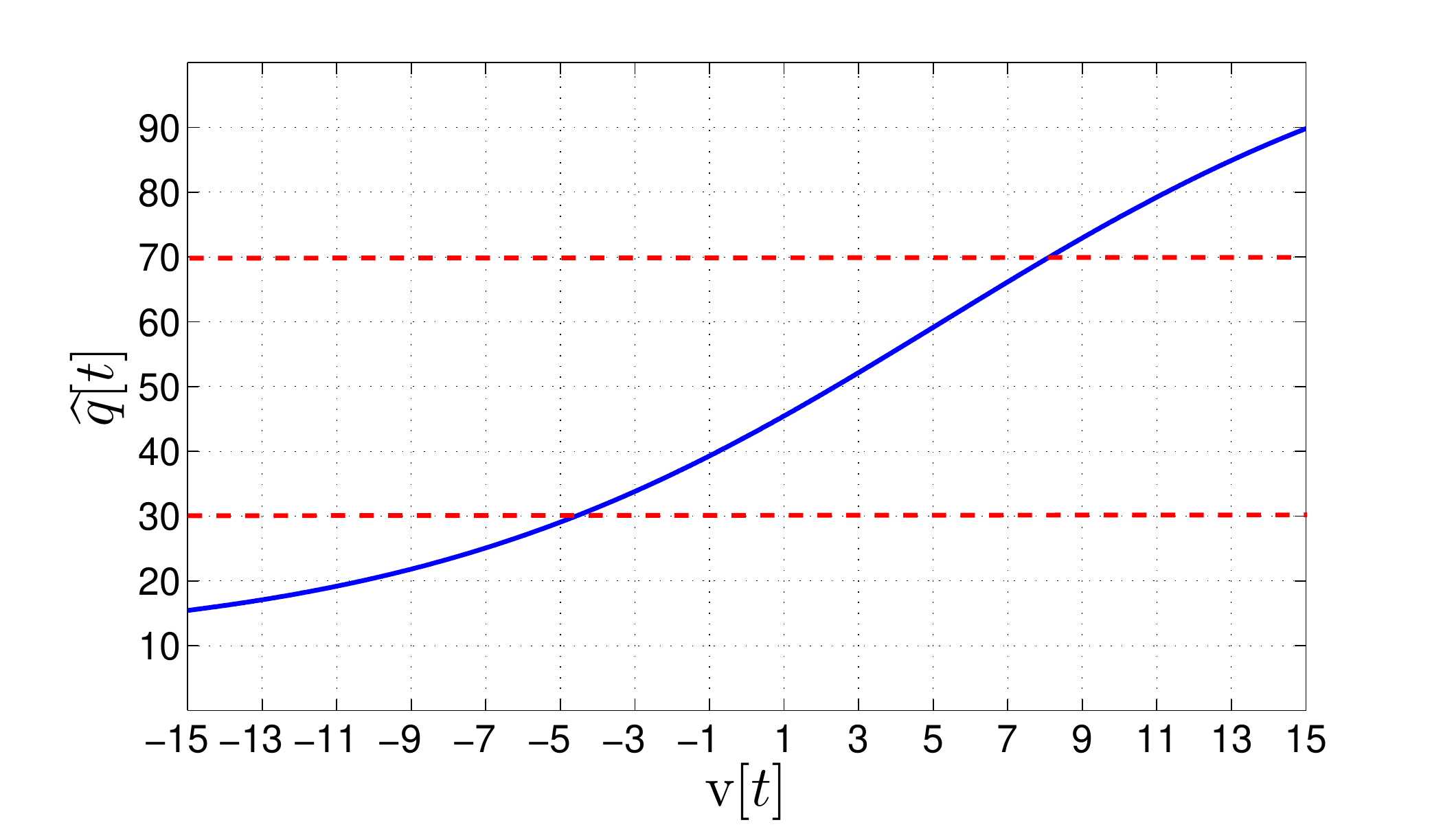}
\label{fig:output}
}
\caption[Optional caption for list of figures]{
Input and output nonlinearities of the HW model.
}
\label{fig:inout}
\end{figure}

In Fig.~\ref{fig:inout}\subref{fig:input}, the input nonlinearity of the TVSQ model is plotted. As the input $\mathrm{q^{st}}[t]$ increases, the gradient of the input nonlinearity diminishes. In particular, the slope of the input nonlinearity is much larger when $\mathrm{q^{st}}[t]<50$. As discussed in Section~\ref{sec:database}, an RDMOS of 50 indicates acceptable STSQ. Therefore, the concavity of the input non-linearity implies that, the TVSQ is more sensitive to quality variations when viewers are watching low quality videos. This also explains why the TVSQ is more sensitive in region C than region D in Fig.~\ref{fig:pre_obs} (see section \ref{sec:pre_obs}).

In Fig.~\ref{fig:inout}\subref{fig:output}, the output nonlinearity of our TVSQ model is plotted. It can be observed that, when $30\leq\widehat{\mathrm{q}}[t]\leq70$, the function is almost linearly increasing with the input. This observation inspired us to further simplify the model by replacing the sigmoid output nonlinearity function with a linear function. Table~\ref{tab:linear_output} shows the performance of the model when the output nonlinearity is replaced by
\begin{equation}
\label{eq:linear_out}
\widehat{\mathrm{q}}[t]=a\mathrm{v}[t]+b,
\end{equation}
where $a=0.7013$ and $b=49.9794$. Comparing with Table~\ref{tab:val}, it can be seen that the outage rate is increased slightly but that the linear correlation coefficients and Spearman's rank correlation coefficients are almost the same. Hence, the simplified model can also predict TVSQ reasonably well. An important advantage of this simplified model is its concavity. Indeed, since the input nonlinearity function is a concave function and the filter is linear, then at any time $t$, the mapping between $\mathrm{q^{st}}[t]$ and $\widehat{\mathrm{q}}[t]$ is also concave. Hence, the simplified model can thus be easily incorporated into a convex TVSQ optimization problem, which can be easily solved and analyzed. 
\begin{center}
\begin{table*}[ht]
\caption{Performance of the model if output nonlinearity is replaced with a linear function}
\centering
\resizebox{\columnwidth}{!}{%
\begin{tabular}{c||l|l|l|l|l|l|l|l|l|l|l|l|l|l|l|l}
    \hline
    &\#1&\#2&\#3&\#4&\#5&\#6&\#7&\#8&\#9&\#10&\#11&\#12&\#13&\#14&\#15&mean\\
    \hline
    outage rate(\%)&12.00 &10.91 &10.55 &16.00 &9.82 &5.45 & 10.18& 11.64&10.55 &10.18 & 4.00&10.91 & 1.45&6.91 &1.09 &{\bf 8.78}\\
    \hline
    linear correlation&0.840 & 0.896& 0.864& 0.787& 0.920& 0.930& 0.869& 0.876& 0.854& 0.842& 0.937& 0.886& 0.883& 0.914& 0.897&{\bf 0.879}\\
    \hline
    rank correlation&0.866&0.845 &0.876 & 0.818&0.906 & 0.939& 0.883& 0.887& 0.851& 0.840& 0.916& 0.890& 0.853& 0.935& 0.853&{\bf 0.877}\\
    \hline
\end{tabular}%
}
\label{tab:linear_output}
\end{table*}
\end{center}
\subsection{Impulse Response of the IIR Filter}
\begin{figure}[h!]
\centering
\includegraphics[width=0.9\columnwidth]{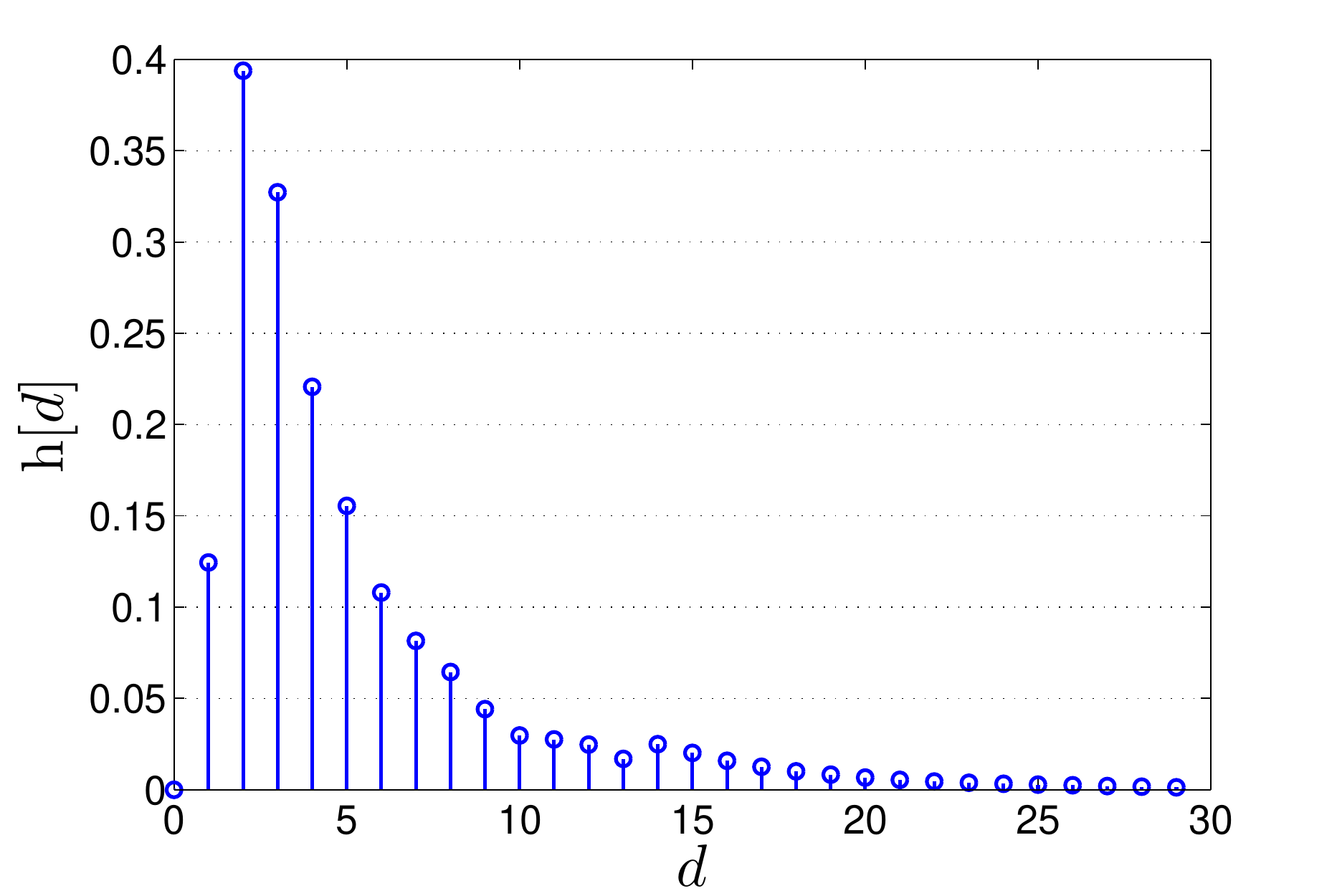}
\caption{The impulse response of the IIR filter in the first 30 seconds.}
\label{fig:impulse}
\end{figure}

The impulse response of the IIR filter in the simplified Hammerstein-Wiener model is shown in Fig.~\ref{fig:impulse}. Denoting the impulse response by $\mathrm{h}[d]$, we have $\mathrm{v}[t]=\sum_{d=0}^\infty\mathrm{h}[d]\mathrm{u}[t-d]$. Thus, $\mathrm{h}[d]$ indicates to what extent the current TVSQ depends on the STSQ of the $d$ seconds prior to the current time. In Fig.~\ref{fig:impulse}, it can be seen that $\mathrm{h}[d]$ is maximized at $d=2$. This means that there is a 2 seconds delay before the viewers respond to a variation in STSQ. That is a natural physiological delay, or latency, between a human subject's observation of STSQ variations and her/his manual response that is given via the human interface. Fig.~\ref{fig:impulse} also shows that $\mathrm{h}[d]$ takes very small values when $d\geq 15$. This implies that the current TVSQ value depends mainly on the STSQs over the immediately preceding 15 seconds. In other words, the visual memory of TVSQ perception on the videos in our database is around 15 seconds. This observation coincides with our analysis that the impact of the initial states of the IIR filter persists for about 15 seconds.

\section{Conclusions and Future Work}
\label{sec:conclusion}
In this paper, a TVSQ prediction model is proposed for the rate-adaptive videos transmitted over HTTP. The model was trained and validated on a new database of quality-varying videos that simulate the true rate-adaptive videos commonly encountered in HTTP-based streaming. Two important conclusions are drawn based on our model. First, the behavioral response of viewers to quality variation is more sensitive in the low quality region than in the high quality region. Second, the current TVSQ can affect the TVSQ in the next 15 seconds. Based on our analysis of the proposed model, the mapping from STSQ and TVSQ is not only monotone but also concave. This property is desirable in solving TVSQ optimization problems.

The proposed TVSQ model can be used to characterize the mapping between video data rate and TVSQ. The rate-adaptation algorithm can then use the rate-TVSQ mapping to select an optimal video data rate that not only avoids playback interruptions but also maximizes the TVSQ.

In this paper, we focus on modeling the impact of quality fluctuations on TVSQ. Of course, frame freezes and re-buffering events caused by playback interruptions can also significantly affect the viewer's QoE. These events, however, are quite distinctive in their source and effect on QoE relative to the types of distortions studied herein. Studying the impact of playback interruptions on TVSQ is an important future work, but is certainly beyond the scope of the work presented here.
\appendices
\section{Instructions for SSCQE}
\label{app:instructions}
You are taking part in a study to assess the quality of videos. You will be shown a video at the center of the monitor and there will be a rating bar at the bottom, which can be controlled by a mouse on the table. You are to provide feedback on how satisfied you are with your viewing experience up to and including the current moment, i.e., by moving the rating bar in real time based on your satisfaction. The extreme right on the bar is `excellent' and the extreme left is `bad'. There is no right or wrong answer.

\section{Gradient Calculation for Model Identification}
\label{app:gradient}
For the parameter $\boldsymbol\gamma$, we have $\nabla_{\boldsymbol\gamma}\mathrm{q^{tv}}[t]=\left(\frac{\partial\mathrm{q^{tv}}[t]}{\partial\gamma_1},\frac{\partial\mathrm{q^{tv}}[t]}{\partial\gamma_2},\frac{\partial\mathrm{q^{tv}}[t]}{\partial\gamma_3},\frac{\partial\mathrm{q^{tv}}[t]}{\partial\gamma_4}\right)^\mathsf{T}$,
where
\begin{equation}
\label{eq:gradient_gamma}
\begin{alignedat}{2}
\frac{\partial\mathrm{q^{tv}}[t]}{\partial\gamma_1}&=\frac{\gamma_4\mathrm{v}[t]\exp(-(\gamma_1\mathrm{v}[t]+\gamma_2)) }{\left(1+\exp(-(\gamma_1\mathrm{v}[t]+\gamma_2))\right)^2},\\
\frac{\partial\mathrm{q^{tv}}[t]}{\partial\gamma_2}&=\frac{\gamma_4\exp(-(\gamma_1\mathrm{v}[t]+\gamma_2)) }{\left(1+\exp(-(\beta_1\mathrm{v}[t]+\beta_2))\right)^2},\\
\frac{\partial\mathrm{q^{tv}}[t]}{\partial\gamma_3}&=1,\\
\frac{\partial\mathrm{q^{tv}}[t]}{\partial\gamma_4}&=\frac{1}{1+\exp\left(-(\gamma_1\mathrm{v}[t]+\gamma_2)\right)}.
\end{alignedat}
\end{equation}
For the parameter $\mathbf b$, $\mathbf f$ and $\boldsymbol\beta$, we have
\begin{align}
\nabla_{\boldsymbol\xi}\mathrm{q^{tv}}[t]=\frac{\partial\mathrm{q^{tv}}[t]}{\partial{\mathrm v}[t]}\nabla_{\boldsymbol\xi}{\mathrm v}[t]=\frac{\gamma_1\gamma_4\exp(-(\gamma_1\mathrm{v}[t]+\gamma_2))}{(1+\exp(-(\gamma_1\mathrm{v}[t]+\gamma_2)))^2}\nabla_{\boldsymbol\xi}{\mathrm v}[t],
\end{align}
where $\boldsymbol\xi$ can be $\mathbf b$, $\mathbf f$ or $\boldsymbol\beta$.
Thus we only need to compute $\nabla_{\boldsymbol\xi}{\mathrm v}[t]$. For $\mathbf b$ and $\mathbf f$, we have
\begin{equation}
\label{eq:gradient_bf}
\begin{alignedat}{2}
\nabla_{\mathbf b}{\mathrm v}[t]&=\left(\mathrm u[t]\right)_{t-1:t-r}+\sum_{d=1}^rf_d\nabla_{\mathbf b}{\mathrm v}[t-d]\\
\nabla_{\mathbf f}{\mathrm v}[t]&=\left(\mathrm v[t]\right)_{t-1:t-r}+\sum_{d=1}^rf_{d}\nabla_{\mathbf f}{\mathrm v}[t-d].
\end{alignedat}
\end{equation}
For $\boldsymbol\beta$, we have
\begin{align}
\label{eq:gradient_beta}
\nabla_{\boldsymbol\beta}{\mathrm v}[t]=\sum_{d=0}^rb_d\nabla_{\boldsymbol\beta}\mathrm{u}[t-{d}]+\sum_{d=1}^rf_{d}\nabla_{\boldsymbol\beta}{\mathrm v}[t-d],
\end{align}
where $\nabla_{\boldsymbol\beta}\mathrm{u}[t]=\left(\frac{\partial\mathrm{u}[t]}{\partial\beta_1},\frac{\partial\mathrm{u}[t]}{\partial\beta_2},\frac{\partial\mathrm{u}[t]}{\partial\beta_3},\frac{\partial\mathrm{u}[t]}{\partial\beta_4}\right)^\mathsf{T}$ can be computed similarly as \eqref{eq:gradient_gamma}.

It may be seen from \eqref{eq:gradient_bf} and \eqref{eq:gradient_beta} that $\nabla_{\mathbf b}{\mathrm v}[t]$, $\nabla_{\mathbf f}{\mathrm v}[t]$ and $\nabla_{\boldsymbol\beta}{\mathrm v}[t]$ can be recursively computed. The stability of the recursions can be ensured by the following lemma.
\begin{lemma}[Stability of recursive gradient calculation]
If the roots of polynomial $1-\sum_{d}^rf_{d}z^{-d}$ are confined within the unit circle of the complex plane, the recursive gradient calculation is stable.
\end{lemma}
\begin{IEEEproof}
In \eqref{eq:gradient_bf} and \eqref{eq:gradient_beta}, the gradients of $\nabla_{\mathbf b}{\mathrm v}[t]$, $\nabla_{\mathbf f}{\mathrm v}[t]$ and $\nabla_{\boldsymbol\beta}{\mathrm v}[t]$ are actually the outputs of IIR filters, where the denominator of the transfer function is $1-\sum_d^rf_dz^{-d}$. Thus the roots of $1-\sum_d^rf_dz^{-d}$ determines the stability of the recursive calculation process and the lemma is proved.
\end{IEEEproof}
To calculate the gradient using \eqref{eq:gradient_rec}, we also need to know the initial values of $\big(\widehat{\mathrm{q}}(t,\boldsymbol\theta)\big)_{1:r}$ to calculate $\left(\frac{\partial\widehat{\mathrm{q}}(t,\boldsymbol\theta)}{\partial\theta_\mathrm{i}}\right)_{1:r}$. For the purpose of model training, we simply set $\big(\widehat{\mathrm{q}}(t,\boldsymbol\theta)\big)_{1:r}=\big(\mathrm{q^{tv}}(t,\boldsymbol\theta)\big)_{1:r}$. Thus, we have $\frac{\partial\widehat{\mathrm{q}}(t,\boldsymbol\theta)}{\partial\theta_\mathrm{i}}=\frac{\partial{\mathrm{q^{tv}}}[t]}{\partial\theta_\mathrm{i}}=0, \forall\,t\leq r$.
\ifCLASSOPTIONcaptionsoff
  \newpage
\fi
\bibliographystyle{IEEEtran}
\bibliography{IEEEabrv,refs,strings}
\end{document}